\documentclass[a4paper]{article}

\usepackage[pdftex]{graphicx}

\usepackage{graphicx}%
\usepackage{multirow}%
\usepackage{amsmath,amssymb,amsfonts}%
\usepackage{amsthm}%
\usepackage{mathrsfs}%
\usepackage[title]{appendix}%
\usepackage{xcolor}%
\usepackage{textcomp}%
\usepackage{manyfoot}%
\usepackage{booktabs}%
\usepackage{algorithm}%
\usepackage{algorithmicx}%
\usepackage{algpseudocode}%
\usepackage{listings}%

\usepackage{bm}
\usepackage{color}
\usepackage{url}

\newtheorem{theorem}{Theorem}
\newtheorem{lemma}{Lemma}
\newtheorem{corollary}{Corollary}
\newtheorem{proposition}{Proposition}
\newtheorem{assumption}{Assumption}
\newcommand{\Imat}{\bm{I}}
\newcommand{\breal}{\mathbb{R}}

\newcommand{\prob}{\mathbf{P}}
\newcommand{\ave}{\mathbf{E}}

\newcommand{\sign}{\mathrm{sign}}
\newcommand{\diag}{\mathrm{diag}}
\newcommand{\probbayes}{p}

\newcommand{\tr}{\mathrm{tr}}

\newcommand{\fracp}[2]{\frac{\partial#1}{\partial#2}}

\newcommand{\fracpps}[2]{\frac{\partial^{2}#1}{\partial#2^{2}}}
\newcommand{\fracppv}[2]{\frac{\partial^{2}#1}{\partial#2\:\partial#2^{\top}}}
\newcommand{\diff}{\: \mathrm{d}}
\newcommand{\argmax}{{\rm \arg} \mathop{\rm max}\limits}
\newcommand{\argmin}{{\rm \arg} \mathop{\rm min}\limits}

\newcommand{\numberset}[2]{\{ #1 \,,\, \ldots \,,\, #2 \}}

\newcommand{\indexvec}[3]{( #1_{#2} \,,\, \ldots \,,\, #1_{#3} )^{\top}}
\newcommand{\iid}{\stackrel{i.i.d.}{\sim}}

\newcommand{\ldts}{\,,\,\ldots\,,\,}

\newcommand{\lconv}{\;\stackrel{\mathrm{L}}{\longrightarrow}\;}

\newcommand{\iniest}{\mathrm{ini}} 
\newcommand{\normaldist}{\mathrm{N}} 

\newcommand{\pbs}{\!\!}
\newcommand{\bseq}{\pbs = \pbs}

\newcommand{\KL}{\mathrm{KL}}
\newcommand{\BHHJ}{\mathrm{BHHJ}}
\newcommand{\JHHB}{\mathrm{JHHB}}
\newcommand{\Cdiv}{\mathrm{C}}
\newcommand{\Criterion}{\mathrm{Criterion}}

\newcommand{\DBBC}{\mathrm{DBBC}}
\newcommand{\EDBBC}{\mathrm{E \mathchar`- DBBC}}
\newcommand{\GEDBBC}{\mathrm{GE \mathchar`- DBBC}}

\title{\large Robust and consistent model evaluation criteria in high-dimensional regression}
\author{Sumito Kurata$^{1}$\thanks{Corresponding author (e-mail: kurata@imi.kyushu-u.ac.jp)} and Kei Hirose$^{1}$ \\
 \tiny $\quad^{1}$ Institute of Mathematics for Industry, Kyushu University, 744 Motooka, Nishi-ku, 819-0395, Fukuoka, Japan. \\
}
\date{}

\begin{document}

\maketitle

\begin{abstract}
Most of the regularization methods such as the LASSO have one (or more) regularization parameter(s), and to select the value of the regularization parameter is essentially equal to select a model. 
Thus, to obtain a model suitable for the data and phenomenon, we need to determine an adequate value of the regularization parameter. 
Regarding the determination of the regularization parameter in the linear regression model, we often apply the information criteria like the AIC and BIC, however, it has been pointed out that these criteria are sensitive to outliers and tend not to perform well in high-dimensional settings. 
Outliers generally have a negative effect on not only estimation but also model selection, consequently, it is important to employ a selection method with robustness against outliers. 
In addition, when the number of explanatory variables is quite large, most conventional criteria are prone to select unnecessary explanatory variables. 
In this paper, we propose model evaluation criteria based on the statistical divergence with excellence in robustness in both of parametric estimation and model selection, by applying the quasi-Bayesian procedure. 
Our proposed criteria achieve the selection consistency even in high-dimensional settings due to precise approximation, simultaneously with robustness. 
We also investigate the conditions for establishing robustness and consistency, and provide an appropriate example of the divergence and penalty term that can achieve the desirable properties. 
We finally report the results of some numerical examples to verify that the proposed criteria perform robust and consistent variable selection compared with the conventional selection methods. 
\end{abstract}

\clearpage

\section{Introduction}
\label{sec:intro}

In the regression analysis, variable selection is essential to understand the cause of a response and background of a phenomenon adequately. 
Information criteria such as the AIC (Akaike's information criterion; \cite{akaike1974new}) and BIC (Bayesian information criterion; \cite{schwarz1978estimating}) are typical methods of the variable selection, and many studies have modified and extended these criteria (e.g., \cite{fujikoshi1997modified, konishi2004bayesian, konishi1996generalised, sugiura1978further}). 
For $n$ observations, let us consider the normal linear regression model with $P$ explanatory variables. 
Suppose that the response variables $Y_{1} \ldts Y_{n}$ are drawn from 
\begin{eqnarray}
 Y_{i} \:=\: \bm{x}_{i}^{\top} \bm{\beta}_{\star} + \epsilon_{i} \:,\quad \epsilon_{i} \iid \normaldist \left( 0 \,,\, \sigma^{2} \right) \quad \left( i \,=\, 1 \ldts n \right) \:, \label{eq:NormalLinearReg}
\end{eqnarray}
where $\bm{x}_{i}$ $=$ $( x_{i, 1} \ldts x_{i, P} )^{\top}$ is the explanatory variable vector of $i$ th individual, $\bm{\beta}_{\star}$ $=$ $\indexvec{\beta^{\star}}{1}{P}$ is the ``true'' value of $P$-dimensional regression coefficient vector $\bm{\beta}$ $=$ $\indexvec{\beta}{1}{P}$, and $\bm{\epsilon}$ $=$ $\indexvec{\epsilon}{1}{n}$ is the error vector consisting of $n$ independent random variables following the normal distribution with zero mean and variance $\sigma^{2} > 0$. 
Let $\bm{Y}$ $=$ $\indexvec{Y}{1}{n}$ be the $n$-dimensional vector consisting of response variables, $\bm{X}$ $=$ $( \bm{x}_{1} \ldts \bm{x}_{n} )^{\top}$ be the $n$ $\times$ $P$ matrix consisting of fixed values of explanatory variables. 
Under the assumption of normality, the minimization of the residual sum of squares, $\| \bm{Y} - \bm{X} \bm{\beta} \|_{2}^{2}$, with respect to $\bm{\beta}$ $\in$ $\breal^{P}$ is equivalent to the of the maximization of the likelihood and the minimization of the Kullback-Leibler divergence (KL divergence; \cite{kullback1951information}), where $\| \cdot \|^{2}$ indicates the $L^{2}$-norm. 
Generally, variable selection methods in the normal linear regression model such as the AIC and BIC assume that every error term (and thus every response variable) is normally distributed, and that the number of explanatory variables is a (not large) constant. 
However, there unfortunately exist quite a few situations that these assumptions are violated in real data analyses. 
In this paper, we focus on two critical causes that degrade the performance of variable selection methods: the {\em outliers} (distinctive, unusual, or mistaken data that are distant from other data) and the {\em high-dimensional} setting. 
The main purpose of this paper is to establish the selection method that solves the two challenges simultaneously, i.e., to propose accurate variable selection method with {\em robustness} against outliers even in the high-dimensional assumptions. 

Firstly, it has been pointed out that, analytical methods built upon the KL divergence such as the parametric estimation and statistical test tend to be sensitive against outliers (e.g., \cite{basu2017wald, ghosh2013robust, kurata2019discrete}). 
Since providing a clear threshold of the outliers or preventing their occurrence are effectively impossible, it is desirable to conduct robust analysis, regardless of whether there exist outliers in the observation set or not. 
Especially, in the model selection problem, information criteria that have the log-likelihood as the main term have a tendency to reduce the accuracy when there exist observations that give extremely small values of the probability density function (or probability function) of model's distribution (see, e.g., \cite{kurata2020consistency, mantalos2010improved}). 
Just as the arithmetic mean is greatly influenced by an extraordinary observation, model evaluation criteria for variable selection without robustness tend to overlook necessary explanatory variables and select unnecessary variables, because of the existence of outliers. 
To solve this problem, statistical divergence measures with robustness against outliers have been applied; \cite{kurata2018robust} derived an AIC-type criterion built upon the {\em Basu-Harris-Hjort-Jones (BHHJ) divergence}, a representative family of divergence measures with robustness against outliers (\cite{basu1998robust, ghosh2013robust}), and \cite{kurata2024robustness} proposed the BIC-type criteria built upon wide classes of robust divergence measures including the BHHJ divergence. 
Recent studies (e.g., \cite{kurata2018robust, kurata2024robustness}) have shown that such criteria have a feature to diminish the negative effect of outliers. 
When discussing the robustness of an analytical method, we often assume that the observations are drawn from a mixture distribution: most observations are drawn from the ``true'' probability distribution, while the minority of the data comes from ``another'' distribution. 
Our goal is to model the ``true distribution'' adequately, and we regard the ``another distribution'' as the source of outliers. 
To assess the robustness of a model evaluation criterion against outliers, we evaluate the difference of the behavior of the criterion between when the mixture proportion of the ``another distribution'' is zero (corresponding to non-contaminated data) and when it is not (contaminated data). 
If the difference becomes (unboundedly) larger depending on the value of outliers, we can see that the corresponding criterion is not robust: this procedure is an extension of the idea of the influence function of an estimator to the model selection problem (refer also to \cite{huber1964robust, hampel1986robust}). 
In other words, robust selection methods tend to maintain the selection results regardless the outliers, and this advantage can be expected to apply to the high-dimensional regression problem. 
Here, we set the first aim of this paper: (i) to perform robust estimation and variable selection that diminish the negative effect caused by the contamination of outliers. 
For achieving this aim, to utilize appropriate class of divergence with desirable robustness is absolutely essential. 

Secondly, when investigating the asymptotic properties of an estimator and model evaluation criterion, most of analytical methods assume that the sample size $n$ is sufficiently large and the number of candidate explanatory variables $P$ is in constant-order, i.e., $P$ $=$ $O(1)$ with respect to $n$, however, we sometimes face to situations where $P$ is comparable with or larger than $n$ in real data analyses. 
In such high-dimensional problems, we often assume the sparsity, that is, the true number of necessary explanatory variables (denoted as $s$ throughout this paper) is sufficiently smaller than $n$ and $P$. 
In recent years, penalized least squares estimation via $L^{1}$-norm-based penalty terms that works well even when $n$ $<$ $P$ have been developed and applied to wide fields. 
The {\em LASSO} (least absolute shrinkage and selection operator; \cite{tibshirani1996regression}) is a typical regularization method that conducts estimation and variable selection simultaneously, by shrinking the estimates of the regression coefficients corresponding to the unnecessary variables to zero. 
Most of the regularization methods including the LASSO have one (or more) regularization parameter(s) that control the intensity of the shrinkage for the values of the regression coefficients, and the variable selection problem is generally equivalent to determine the regularization parameter. 
As we do not know the true subset of explanatory variables in advance, the {\em oracle property}, the ability to identify the true subset as if it was given, is desirable in variable selection. 
\cite{fan2001variable} developed a non-concave penalty function, named {\em SCAD} (smoothly clipped absolute deviation) penalty, and showed the superiority of the SCAD over the original LASSO in terms of the oracle property. 
\cite{zou2006adaptive} alternatively introduced the {\em adaptive LASSO}, a generalization of the original LASSO, by assigning different weights to different coefficients, and proved their estimation procedure achieved the oracle property. 
However, it should be noted that the oracle property provides us with some asymptotic orders of the regularization parameter, yet it does not provide one ``optimal value'': generally, any constant multiple of a value achieving the oracle property also achieves the oracle property. 
Thus, we need an optimal value of the regularization parameter to conduct reliable variable selection, because, in practical situations, we unfortunately do not know the true subset of explanatory variables in advance. 
Most of commonly used criteria for determining the optimal value of the regularization parameter such as the AIC and BIC have a drawback in addition to non-robustness against outliers: it is lack of the {\em selection consistency} in high-dimensional settings. 
The selection consistency, the property that the probability of selecting the ``true model'' in the candidates tends to one when the sample size $n$ diverges infinity, is an important asymptotic property in model selection. 
Although the BIC has the selection consistency in the case of fixed dimensionality ($P$ $=$ $O(1)$), it has been pointed out that the original BIC can not keep the selection consistency when $P$ $=$ $P_{n}$ also goes to infinity (e.g., \cite{chen2008extended, chen2012extended}). 
Therefore, for detecting the necessary explanatory variables adequately, we set the second aim: (ii) to establish the selection consistency for model evaluation criteria build upon statistical divergence under the high-dimensional assumptions. 

In this paper, we propose the model evaluation criteria that accomplish the above-mentioned two aims (i) and (ii) for determing the regularization parameter of the LASSO-type regularization methods: as per the previous paragraphs, there are some previous researches that have achieved one either of the robustness or consistency in selection, but we attempt to realize them simultaneously. 
For this purpose, we introduce the penalized (quasi) likelihood built upon robust divergence measures, and approximate the (quasi) posterior probability based on the divergence, by generalizing the deriving process of the BIC-type criteria. 
For achieving the aim (ii), it is necessary to approximate the (quasi) posterior probability of each model precisely, considering the relationship among $n$, $P$, and $s$. 
In addition, to obtain desirable properties of the sparse estimators and to derive accurate criteria from divergence measures, we need to employ appropriate penalty term of the regularization method. 
We present theorems for achieving the two aims, and provide some specific divergence and penalty term fulfilling the conditions for the theorems. 
Our proposed criteria possess a characteristic of reducing the negative effect of outliers by down-weighting for them, and simultaneously, the proposed criteria achieve the selection consistency even in high-dimensional settings. 
The robustness in variable selection and selection consistency hold not only in the case of fixed dimensionality but also in high-dimensional settings. 

This paper is organized as follows. 
In Section \ref{sec:GALASSO}, we first review the LASSO and related methods, and some classes of statistical divergence measures with robustness against outliers, then attempt to generalize the idea of the LASSO going along with the robust divergence. 
In Section \ref{sec:criteria}, we derive model evaluation criteria by utilizing a quasi-Bayesian procedure based on divergence measures and sparse regularization methods, and discuss what divergence and penalty term can achieve the two aims (i) and (ii). 
We verify the performance of the proposed method by several numerical simulations and real data analysis in Section \ref{sec:numerical}. 
We finally conclude this paper in Section \ref{sec:conc}. 
The regularity conditions, proofs of theorems, and additional results of numerical experiments are presented in the Appendix.

\section{The LASSO-type regularization and its generalization}
\label{sec:GALASSO}

For the normal linear regression model with sample size $n$ (given in Eq.~\eqref{eq:NormalLinearReg}), we hereafter assume that the number of candidate explanatory variables $P$ $=$ $P_{n}$ is proportional to $\exp (n^{l})$ for some $l$ $\in$ $(0 ,\, 1)$. 
The relationship between $n$ and $P$ has been assumed in many studies about the high-dimensional problem (e.g., \cite{basu2024robust, chen2012extended, fan2014adaptive, huang2008adaptive}), and this assumption covers both of the case of $n$ $>$ $P$ (when $l$ is close to $0$) and $n$ $<$ $P$ (when $l$ is distant from $0$). 
In this paper, we discuss the estimation of $\bm{\beta}$ and detection of the necessary explanatory variables, and we regard the variance of the error terms $\sigma^{2}$ as a nuisance parameter. 
We assume that all elements in the ``true'' regression coefficient vector $\bm{\beta}_{\star}$ are $O(1)$ (constant-order with respect to $n$), and $\bm{\beta}_{\star}$ has $s$ non-zero elements. 
We suppose that $s$ $=$ $s_{n}$, corresponding to the number of necessary explanatory variables, diverges to infinity relatively slowly than $n$ and $P$. 
This assumption corresponds to the purpose of detecting (relatively) few ``truly'' necessary variables in numerous candidates: this is often supposed in various fields, for example, the finance, bioinformatics, genetics, and computer science. 

We now consider the estimator $\hat{\bm{\beta}}$ of $\bm{\beta}$, obtained by the following minimizing problem: 
\begin{eqnarray}
 \hat{\bm{\beta}} \:=\: \argmin_{\bm{\beta} \in \breal^{P}} \left\{ \sum_{i=1}^{n} \, \rho_{i}(Y_{i} - \bm{x}_{i}^{\top} \bm{\beta}) \,+\, n \, \lambda \, \left\| \bm{w} \circ \bm{\beta} \right\|_{1} \right\} \:, \label{eq:GALASSO:estimator}
\end{eqnarray}
where $\lambda$ $=$ $\lambda_{n}$ is a non-negative regularization parameter, $\bm{w}$ $=$ $\indexvec{w}{1}{P}$ is the $P$-dimensional weight vector consisting of non-negative elements, $\| \cdot \|_{1}$ means the $L^{1}$-norm, and $\circ$ indicates the Hadamard product (element-wise product). 
We assume that, the main term (the first term in Eq.~\eqref{eq:GALASSO:estimator}) is composed of non-negative-valued functions $\rho_{1} \ldts \rho_{n}$, and these functions are convex in $\Theta_{\star}$, an open set containing $\bm{\beta}_{\star}$. 
The minimization problem given in Eq.~\eqref{eq:GALASSO:estimator} is a generalization of the LASSO (\cite{tibshirani1996regression}) and the adaptive LASSO (\cite{zou2006adaptive}): if $\rho_{i}(z)$ $=$ $z^{2} / 2$ for all $i$ $=$ $1 \ldts n$, the minimizer of Eq.~\eqref{eq:GALASSO:estimator} coincides with the adaptive LASSO estimates. 
We can see that the (original) LASSO is the special case where the weight vector $\bm{w}$ is the $P$-dimensional unit vector $\bm{1}_{P}$ $=$ $\left( 1 \ldts 1 \right)^{\top}$. 
Generally, every value of the element in $\bm{w}$ is given preliminarily: in many studies, it depends on an {\em initial estimator} $\hat{\bm{\beta}}^{\iniest}$ $=$ $\indexvec{\hat{\beta}^{\iniest}}{1}{P}$. 
As the initial estimator, the ordinary least squares (OLS) estimates, ridge estimates, and (non-adaptive) LASSO-type estimates have been often used (refer also to, e.g., \cite{ghosh2024robust, schneider2012catching, wagener2012bridge}). 
In this paper, we assume that each weight $w_{j}$ has the form of $w_{j}$ $=$ $h(|\hat{\beta}^{\iniest}_{j}|)$ for each $j$ $=$ $1 \ldts P$ and some non-increasing function (we refer to it as {\em weight function}) $h(\cdot)$ over $(0 ,\, +\infty)$.

\subsection{A generalization of the LASSO via robust statistical divergence}
\label{sec:GALASSO:divergence}

For the LASSO-type regularization, we usually employ the residual sum of squares (RSS) as the main term in Eq.~\eqref{eq:GALASSO:estimator}. 
Minimizing of the RSS is equivalent to estimating the regression coefficients based on the KL divergence when the error terms follow the normal distribution, however, it has been pointed that the KL divergence-based methods are not robust against outliers. 
Therefore, in this paper, we consider applications of statistical divergence measures with robustness to the main term. 
For each $i$ $=$ $1 \ldts n$, let $D_{i}$ be a divergence measure between the ``true'' probability distribution of $Y_{i}$ (in the case of the normal linear regression, it corresponds to $\normaldist(\bm{x}_{i}^{\top} \bm{\beta}_{\star} ,\, \sigma^{2})$, the normal distribution with mean $\bm{x}_{i}^{\top} \bm{\beta}_{\star}$ and variance $\sigma^{2}$) and the distribution of a statistical model (in our case, it corresponds to $\normaldist(\bm{x}_{i}^{\top} \bm{\beta} ,\, \sigma^{2})$). 
We here introduce two classes of divergence measures excellent in robustness against outliers, {\em JHHB divergence family} and {\em C divergence family}. 

For each response variable $Y_{i}$, let $g_{i}(\cdot)$ be the probability (density) function of the true distribution and $f_{i}(\cdot\,|\,\bm{\beta})$ be the one of the model (in our case, $g_{i}(\cdot)$ $=$ $f_{i}(\cdot\,|\,\bm{\beta}_{\star})$). 
The JHHB divergence family, proposed by \cite{jones2001comparison}, has the following form: 
\begin{eqnarray}
 D^{\JHHB}_{i}(\bm{\beta}) &\bseq& \frac{1}{\varphi} \left\{ \int f_{i}(y\,|\,\bm{\beta})^{\alpha + 1} \diff y \right\}^{\varphi} - \frac{\alpha + 1}{\varphi \, \alpha} \left\{ \int f_{i}(y\,|\,\bm{\beta})^{\alpha} \, g_{i}(y) \diff y \right\}^{\varphi} \nonumber \\
 && \quad+\, \frac{1}{\varphi \, \alpha} \left\{ \int g_{i}(y)^{\alpha + 1} \diff y \right\}^{\varphi} \quad \left( i \,=\, 1 \ldts n \right) \:, \label{eq:JHHBdiv}
\end{eqnarray}
where $\alpha$ and $\varphi$ are positive tuning parameters associated with this divergence family. 
When $\alpha$ goes to $0$, the JHHB divergence tends to the KL divergence, $D^{\KL}_{i}(\bm{\beta})$ $=$ $\int g_{i}(y) \, \log \frac{g_{i}(y)}{f_{i}(y\,|\,\bm{\beta})} \diff y$. 
This divergence for $\varphi$ $=$ $1$ is known as the BHHJ divergence (also called density power divergence or DPD; proposed by \cite{basu1998robust}), i.e., 
\begin{eqnarray}
 D^{\BHHJ}_{i}(\bm{\beta}) &\bseq& \int f_{i}(y\,|\,\bm{\beta})^{\alpha + 1} \diff y - \frac{\alpha + 1}{\alpha} \int f_{i}(y\,|\,\bm{\beta})^{\alpha} \, g_{i}(y) \diff y  \nonumber \\
 && \quad+\, \frac{1}{\alpha} \int g_{i}(y)^{\alpha + 1} \diff y \quad \left( i \,=\, 1 \ldts n \right) \:. \label{eq:BHHJdiv}
\end{eqnarray}
In addition, when $\varphi$ goes to $0$, it tends to the logarithmic density power divergence (also called the gamma divergence). 
Noting that, \cite{jones2001comparison} pointed out that this family has exact unbiasedness only when $\varphi$ is equal to $1$ or $0$ in parametric estimation. 

The C divergence family is another wide class of divergence measures, derived by \cite{vonta2012properties} and \cite{maji2019robust} independently, that has the following form: 
\begin{eqnarray}
 D^{\Cdiv}_{i}(\bm{\beta}) \:=\: \int N \left( \frac{g_{i}(y)}{f_{i}(y\,|\,\bm{\beta})} - 1 \right) \, f_{i}(y\,|\,\bm{\beta})^{\alpha + 1} \diff y \quad \left( i \,=\, 1 \ldts n \right) \:, \label{eq:Cdiv}
\end{eqnarray}
where $\alpha$ is a positive tuning parameter that controls the robustness as with the one of the JHHB divergence family, and $N$ is a strictly convex function on $[-1, +\infty)$, that is three times continuous differentiable and satisfy $N(0)$ $=$ $0$, $N'(0)$ $=$ $0$, and $N''(0)$ $>$ $0$. 
The JHHB and C divergence families given in Eqs.~\eqref{eq:JHHBdiv} and \eqref{eq:Cdiv} have no inclusion relation, but both of them include the KL divergence and BHHJ divergence: when $N(z)$ $=$ $1$ $-$ $\frac{\left( \alpha + 1 \right) \, \left( z + 1 \right)}{\alpha}$ $+$ $\frac{\left( z + 1 \right)^{\alpha + 1}}{\alpha}$, we can see that Eq.~\eqref{eq:Cdiv} coincides with Eq.~\eqref{eq:JHHBdiv} for $\varphi$ $=$ $1$. 

Based on the above divergence families, we can construct the main term of the LASSO-type regularization (Eq.~\eqref{eq:GALASSO:estimator}) as $\sum_{i=1}^{n}$ $\rho_{i}(Y_{i} - \bm{x}_{i}^{\top} \bm{\beta})$, where $\rho_{i}$ ($i$ $=$ $1 \ldts n$) is obtained by replacing the (unknown) true distribution with the empirical distribution based on the observations, and by subtracting parts independent of the estimating target $\bm{\beta}$ from $D_{i}(\bm{\beta})$. 
For example, $\rho_{i}$ for the BHHJ divergence is calculated as follows: 
\begin{eqnarray}
 \rho^{\BHHJ}_{i}(z) \:=\: \int f_{i}(z'\,|\,\bm{\beta})^{\alpha + 1} \diff z' - \frac{\alpha + 1}{\alpha} \, f_{i}(z\,|\,\bm{\beta})^{\alpha} \quad \left( i \,=\, 1 \ldts n \right) \:. \label{eq:rho:BHHJ}
\end{eqnarray}

In the rest of this paper, we mainly focus on the BHHJ divergence. 
An important cause of the lack of robustness against outlier of the RSS- and KL divergence-based estimation methods is that they treat all observations (residuals) including the outliers equally. 
If there exist some outliers in the observation set, corresponding residuals will take extremely large values, and non-robust methods will be notably affected by them. 
A key characteristic of the BHHJ divergence in parametric estimation is the down-weighting based on the value of the probability (density) function of each observation. 
Outliers that are wildly discrepant with respect to the true model's distribution will be given nearly zero weights. 
The tuning parameter $\alpha$ controls the trade-off between efficiency and robustness in estimation: larger $\alpha$ $>$ $0$ treats outliers severely (refer to \cite{basu1998robust, ghosh2013robust}). 
Many previous studies have reported that analytical methods built upon the BHHJ divergence deliver superior robustness in many problems (refer to, e.g., \cite{ghosh2016robustbayes, ghosh2016robustestimation, kurata2019discrete}), and some studies have shown that the BHHJ divergence contributes not only robust parametric estimation but also in statistical hypothesis tests (e.g., \cite{balakrishnan2019robust, basu2017wald}) and model selection (e.g., \cite{avlogiaris2019criterion, kurata2024robustness, mantalos2010improved}). 
In the field of regularization, \cite{basu2024robust, ghosh2024robust} recently proposed the (adaptive) LASSO-type estimation procedure utilizing the BHHJ divergence and discussed the asymptotic behavior of the estimator. 
In Section \ref{sec:criteria}, we will propose new criteria for the selection of the regularization parameter in the LASSO-type method, that enable robust variable selection even in high-dimensional settings.

\subsection{Asymptotic properties of the adaptive LASSO-type estimators based on divergence measures}
\label{sec:GALASSO:properties}

In this subsection, we introduce the asymptotic properties of the estimator given in Eq.~\eqref{eq:GALASSO:estimator}. 
For the true regression coefficient vector $\bm{\beta}_{\star}$, let define $\mathcal{J}^{(1)}$ and $\mathcal{J}^{(2)}$ as $\{ j \,|\, \beta^{\star}_{j} \neq 0 \}$ and $\{ j \,|\, \beta^{\star}_{j} = 0 \}$, respectively. 
We can see that, to select an true combination of explanatory variables is equivalent to find out the subset $\mathcal{J}^{(1)}$ composed of the indices of the necessary explanatory variables. 
Additionally, for a $P$-dimensional vector $\bm{v}$ $=$ $\indexvec{v}{1}{P}$ and for a $P$-columns matrix $\bm{M}$ $=$ $(\bm{m}_{1} \ldts \bm{m}_{P})$, we define the $s$-dimensional vector $\bm{v}^{(1)}$ and $s$-columns matrix $\bm{M}^{(1)}$ as $(v_{j})_{j \in \mathcal{J}^{(1)}}$ and $(\bm{m}_{j})_{j \in \mathcal{J}^{(1)}}$, respectively, and define the $(P-s)$-dimensional vector $\bm{v}^{(2)}$ and $(P-s)$-columns matrix $\bm{M}^{(2)}$ similarly. 
For the sake of simplicity, in this subsection, we suppose that only the first $s$ elements in $\bm{\beta}_{\star}$ are non-zero, i.e., $\mathcal{J}^{(1)}$ $=$ $\numberset{1}{s}$ and $\mathcal{J}^{(2)}$ $=$ $\numberset{s+1}{P}$. 

We here give some notations relative to the functions $\rho_{1} \ldts \rho_{n}$ in Eq.~\eqref{eq:GALASSO:estimator}. 
Let $\check{\bm{d}}_{n}$ be a random vector consisting of the first-derivatives of them, $\check{\bm{d}}_{n}$ $=$ $( \rho_{1}'(\epsilon_{1}) \ldts \rho_{n}'(\epsilon_{n}) )^{\top}$. 
We assume the expectation vector of $\check{\bm{d}}_{n}$ is zero vector, and denote the covariance matrix of $\check{\bm{d}}_{n}$ by $\bm{\Omega}_{n}$. 
For example, we can easily confirm that $\ave [ \rho_{i}'(\epsilon_{i}) ]$ is equal to zero for each $i$ $=$ $1 \ldts n$ for the BHHJ divergence-based functions $\rho_{1} \ldts \rho_{n}$ (see also Eq.~\eqref{eq:rho:BHHJ}). 
Then, let $\bm{D}_{n}$ $=$ $\diag ( \ave [ \rho_{1}''(\epsilon_{1}) ] \ldts \ave [ \rho_{n}''(\epsilon_{n}) ] )$ be a $n$ $\times$ $n$ (non-random) diagonal matrix consisting of the expectations of the second-derivatives. 
In addition, we define a $s$ $\times$ $s$ matrix $\bm{V}_{n}$ $=$ $( \bm{X}^{(1) \top} \bm{D}_{n} \, \bm{X}^{(1)} )^{-\frac{1}{2}}$ and a $n$ $\times$ $s$ matrix $\bm{Z}_{n}$ $=$ $\bm{X}^{(1)} \, \bm{V}_{n}$. 
Regarding these terms, we give some conditions in order to prove asymptotic properties in Subsection \ref{secappend:conditions:Az}. 

The following proposition is known as the oracle property, and it is a generalization of the result by \cite{ghosh2024robust} along wide classes of divergence measures. 

\begin{proposition}
\label{proposition:estimation}
Under (A-1)--(A-6) in Assumption \ref{assumption:estimation}, there exists an optimizer $\hat{\bm{\beta}}$ $=$ $( \hat{\bm{\beta}}^{(1) \top},$ $\hat{\bm{\beta}}^{(2) \top} )^{\top}$ of Eq.~\eqref{eq:GALASSO:estimator} in a neighborhood of $\bm{\beta}_{\star}$ that satisfies
\begin{eqnarray}
 \hat{\bm{\beta}}^{(2)} \:=\: \bm{0}_{P-s} \:,\quad \bm{U}_{n} \left( \hat{\bm{\beta}}^{(1)} - \bm{\beta}_{\star}^{(1)} + \bm{b}_{n} \right) \lconv \normaldist \left( \bm{0}_{s} \,,\: \Imat_{s \times s} \right) \:, \nonumber
\end{eqnarray}
with probability tending to one as $n$ $\to$ $+\infty$, where $\bm{U}_{n}$ $=$ $( \bm{Z}_{n}^{\top} \bm{\Omega}_{n} \, \bm{Z}_{n} )^{-\frac{1}{2}} \bm{V}_{n}^{-1}$, $\bm{b}_{n}$ $=$ $n \, \lambda \, \bm{V}_{n}^{2} \, \tilde{\bm{w}}^{(1)}$, and $\tilde{\bm{w}}^{(1)}$ $=$ $( \tilde{w}_{j} )_{j}$ indicates a $s$-dimensional vector defined as $\tilde{w}_{j}$ $=$ $w_{j} \, \sign ( \beta^{\star}_{j} )$ for each $j$ $\in$ $\mathcal{J}^{(1)}$. 
\end{proposition}

The consistency of estimator $\hat{\bm{\beta}}^{(2)}$ and the asymptotic normality of $\hat{\bm{\beta}}^{(1)}$ can be proved almost similarly to Theorem 3 in \cite{fan2014adaptive} and Theorem 4.3 in \cite{ghosh2024robust}. 
Recently, \cite{ghosh2024robust} derived that the original LASSO- and adaptive LASSO-type estimation procedures built upon the BHHJ divergence, and showed the influence function of the estimator, a measure for assessing the damage caused by a perturbation in data-generating distribution (see also \cite{hampel1986robust, avella2017influence}), is bounded for arbitrary outliers when we use the minimum BHHJ divergence estimator (unregularized, or with a sufficiently small regularization parameter) as the initial estimator. 
Their results can be generalized for the JHHB and C divergence families due to the boundedness of the influence functions based on the divergence families in the case of the normal distribution (refer also to \cite{jones2001comparison, maji2019robust}).

\section{Model evaluation criteria with robustness and selection consistency in the high-dimensional regression}
\label{sec:criteria}

Proposition \ref{proposition:estimation} in the previous section supports that the estimators based on regularization methods satisfying the assumptions, such as the adaptive LASSO-type regularization methods via statistical divergence, have the desirable asymptotic characteristic of the oracle property. 
Nevertheless, Proposition \ref{proposition:estimation} and related assumptions (Assumption \ref{assumption:estimation} in the Appendix) do not give an optimal value the regularization parameter $\lambda$, but just provide some asymptotic orders of $\lambda$. 

To determine a single optimal value (corresponding a combination of explanatory variables) of the regularization parameter, various information criteria other than the original AIC and BIC have been proposed and applied. 
\cite{shibata1989statistical} introduced the regularization information criterion (RIC) for model selection of the penalized likelihood problems in a similar derivation to the AIC. 
The generalized information criteria (GIC; \cite{konishi1996generalised}) that includes the RIC can also be used for regularization parameter selection. 
\cite{konishi2004bayesian} generalized the BIC to evaluate models estimated by the maximum penalized likelihood or regularization methods: their criteria is referred as to GBIC. 
\cite{chen2008extended, chen2012extended, hui2015tuning} introduced criteria that has the selection consistency when $P$ also grows with $n$. 
Moreover, some criteria for regularization parameter selection when employing sparse regularization methods such as the LASSO have been proposed (e.g., \cite{fan2013tuning, hirose2012variable, ninomiya2016aic, umezu2019aic}). 
Most of these share a common ground that are built upon the residual sum of squares or log-likelihood, thus we can regard that such criteria are based on the KL divergence. 
However, determination methods of the regularization parameter based on the KL divergence mentioned above tend to be greatly influenced by an extraordinary observation. 
We therefore employ robust divergence measures instead of the KL divergence and derive model evaluation criteria probability. 
In this section, we derive BIC-type model evaluation criteria with robustness against outliers and selection consistency even in the high-dimensional setting, by utilizing the quasi-Bayesian procedure. 

Let $\{ \mathcal{M}_{\iota} \}_{\iota}$ be the set of candidate models, $\nu(\iota)$ be the number of explanatory variables employed in model $\mathcal{M}_{\iota}$, and $\bm{\beta}(\iota)$ be the corresponding $\nu(\iota)$-dimensional regression coefficient vector. 
We here denote the ``true model'' and the ``true coefficient vector'' by $\mathcal{M}_{\iota_{\star}}$ and $\bm{\beta}_{\star}$ $=$ $\bm{\beta}_{\star}(\iota_{\star})$ (this is a $\nu(\iota_{\star})$ $=$ $s$-dimensional vector), respectively. 
Since we assume that the number of necessary explanatory variables $s$ is rather smaller than $n$ and $P$ (see also Section \ref{sec:GALASSO}), we do not need to take models having quite large numbers of explanatory variables into account. 
Hereafter, we consider to select a model only from the candidate set $\{ \mathcal{M}_{\iota} \}_{\iota}$ fulfilling $\nu(\iota)$ $\leq$ $S$ for some $S$ $=$ $O(s)$ (obviously, $S$ $\geq$ $s$). 
Now, we regard the following function as the {\em quasi-likelihood} under model $\mathcal{M}_{\iota}$: 
\begin{eqnarray}
 M_{n}(\bm{\beta}(\iota)) \:=\: - \sum_{i=1}^{n} \, \rho_{i} \left( Y_{i} - \bm{x}_{i}(\iota)^{\top} \bm{\beta}(\iota) \right) \:, \label{eq:criteria:Mn}
\end{eqnarray}
where $\bm{x}_{i}(\iota)$ is the $\nu(\iota)$-dimensional explanatory variable vector for $i$ th ($i$ $=$ $1 \ldts n$) observation corresponding to model $\mathcal{M}_{\iota}$. 
By (A-1) in Assumption \ref{assumption:estimation}, $M_{n}(\bm{\beta}(\iota))$ is three times differentiable with respect to $\bm{\beta}(\iota)$. 
We here define a $\nu(\iota)$ $\times$ $\nu(\iota)$ matrix, $\bm{H}_{M_{n}}(\bm{\beta}(\iota))$ $=$ $- \fracppv{M_{n}(\bm{\beta}(\iota))}{\bm{\beta}(\iota)}$. 
Let $\pi_{\iota}$ be the prior density of $\bm{\beta}(\iota)$ under model $\mathcal{M}_{\iota}$, that is proper and twice continuously differentiable, and define the estimator as 
\begin{eqnarray}
 \hat{\bm{\beta}}(\iota) \:=\: \argmax_{\bm{\beta}(\iota)} \, R_{n}(\bm{\beta}(\iota)) \:,\quad R_{n}(\bm{\beta}(\iota)) \:=\: M_{n}(\bm{\beta}(\iota)) + \log \pi_{\iota}(\bm{\beta}(\iota)) \:. \nonumber
\end{eqnarray}
In the normal linear regression models, the maximizer coincides with the adaptive LASSO-type estimator in Eq.~\eqref{eq:GALASSO:estimator} when we use the Laplace distribution (double exponential distribution) whose density function is 
\begin{eqnarray}
 \pi_{\iota}(\bm{\beta}(\iota)) &\bseq& \prod_{j \in \mathcal{J}(\iota)^{(1)}} \frac{n \, \lambda \, w_{j}}{2} \, \exp \left( - n \, \lambda \, w_{j} \left| \beta_{j} \right| \right) \nonumber \\
 &\bseq& \left( \frac{n \, \lambda}{2} \right)^{\nu(\iota)} \left( \prod_{j \in \mathcal{J}(\iota)^{(1)}} w_{j} \right) \, \exp \left( - n \, \lambda \sum_{j \in \mathcal{J}(\iota)^{(1)}} w_{j} \left| \beta_{j} \right| \right)
 \label{eq:criteria:GALASSOprior}
\end{eqnarray}
as the prior distribution, where $\mathcal{J}(\iota)^{(1)}$ is the subset of $\numberset{1}{P}$ consisting of the indices corresponding to explanatory variables employed in model $\mathcal{M}_{\iota}$. 
Then, for a model $\mathcal{M}_{\iota}$ with prior probability $\probbayes(\iota)$, we define the quasi-marginal distribution and quasi-posterior probability with respect to the model $\mathcal{M}_{\iota}$ (refer also to \cite{ghosh2016robustbayes}), as follows: 
\begin{eqnarray}
 m_{\iota}(\bm{Y}) \:=\: \int \exp \left\{ M_{n}(\bm{\beta}(\iota)) \right\} \, \pi_{\iota}(\bm{\beta}(\iota)) \diff \bm{\beta}(\iota) \:,\quad \probbayes(\iota \,|\, \bm{Y}) \:=\: \frac{m_{\iota}(\bm{Y}) \, \probbayes(\iota)}{\sum_{\iota'} m_{\iota'}(\bm{Y}) \, \probbayes(\iota')} \:. \nonumber
\end{eqnarray}
BIC-type criteria can be regarded as the approximations of $(-2)$ times of the logarithm of the numerator of the quasi-posterior probability $\probbayes(\iota \,|\, \bm{Y})$, i.e., $- 2 \, \log m_{\iota}(\bm{Y})$ $-$ $2 \, \log \probbayes(\iota)$. 
As can be seen from the formula, the smaller the value of a BIC-type criterion, the ``better'' the corresponding model $\mathcal{M}_{\iota}$. 
When using the log-likelihood as $M_{n}$ (Eq.~\eqref{eq:criteria:Mn}) and supposing that the prior probability $\probbayes(\iota)$ is uniform on the candidate set, we have the original BIC (\cite{schwarz1978estimating}). 
Moreover, we also obtain the {\em DBBC} (divergence-based Bayesian criterion; \cite{kurata2024robustness}) when using $M_{n}$ based on statistical divergence such as the BHHJ and JHHB divergence families. 
The definition of the DBBC-type criterion including the BIC is given as $-2 \, M_{n}(\hat{\bm{\beta}}(\iota))$ $+$ $\nu(\iota) \log n$. 

It is known that the original BIC and some related criteria like the DBBC have the selection consistency in the case of fixed dimensionality, i.e., $P$ $=$ $O(1)$ (e.g., \cite{bozdogan1987model, nishii1984asymptotic}), yet, the selection consistency can not be guaranteed in high-dimensional settings (i.e., $P$ also grows with $n$). 
\cite{chen2008extended} proposed the extended BIC (EBIC) that keep the selection consistency even if $P$ diverges to infinity, by introducing another prior probability of each model $\mathcal{M}_{\iota}$ (denoted as $\probbayes(\iota)$). 
The EBIC and its related criteria (e.g., \cite{chen2012extended, hui2015tuning}) overcame the important drawback of the BIC, selection consistency in high-dimensional settings, nevertheless, such criteria still suffer from another weak point, non-robustness against outliers, because they are based on the log-likelihood and thus the KL divergence. 
Thus, we further extend the EBIC via robust divergence measures, going along with the idea of \cite{chen2008extended, chen2012extended}. 
Suppose that the candidate set is partitioned into $\biguplus_{j=1}^{P} \mathcal{M}_{(j)}$, where $\mathcal{M}_{(j)}$ $=$ $\{ \mathcal{M}_{\iota} \,|\, \nu(\iota) = j \}$ for each $j$ $=$ $1 \ldts P$, i.e., every candidate model in the set $\mathcal{M}_{(j)}$ has $j$ necessary explanatory variables. 
Assume that the conditional probability $\probbayes (\iota \,|\, \mathcal{M}_{(j)})$ is equal to the reciprocal of the cardinality of $\mathcal{M}_{(j)}$, and assign the prior probability $\probbayes(\mathcal{M}_{(j)})$ proportional to $(\# \mathcal{M}_{(j)})^{\gamma}$ for some constant $\gamma$ $\in$ $(0 ,\, 1)$. 
These imply that the prior probability $\probbayes(\iota)$ is proportional to $(\# \mathcal{M}_{(\nu(\iota))})^{-(1 - \gamma)}$ (for more details, refer to \cite{chen2008extended}). 
In the case of the normal linear regression model, $\# \mathcal{M}_{(j)}$ $=$ $1 / \binom{P}{j}$ for each $j$, and $\probbayes(\iota)$ is approximately proportional to $P^{-(1 - \gamma) \, \nu(\iota)}$ for sufficiently large $P$. 
Thus, by applying the Laplace approximation (e.g., \cite{kass1990validity}), we obtain the following approximation of $\log m_{\iota}(\bm{Y})$ $+$ $\log \probbayes(\iota)$: 
\begin{eqnarray}
 && M_{n}(\hat{\bm{\beta}}(\iota)) + \log \pi_{\iota}(\hat{\bm{\beta}}(\iota)) + \frac{\nu(\iota)}{2} \left\{ \log \left( 2 \, \pi \right) - \log n \right\} \nonumber \\
 && \quad-\, \frac{1}{2} \, \log \left| \bm{T}_{n}(\hat{\bm{\beta}}(\iota)) \right| - (1 - \gamma) \, \nu(\iota) \log P \:, \label{eq:criteria:derivation:LapApp}
\end{eqnarray}
in a similar ways as \cite{chen2008extended, hui2015tuning, kurata2024robustness}, where $\bm{T}_{n}(\bm{\beta}(\iota))$ $=$ $-\frac{1}{n} \fracppv{R_{n}(\bm{\beta}(\iota))}{\bm{\beta}(\iota)}$. 
As every element in $\bm{\beta}(\iota)$ is non-zero under model $\mathcal{M}_{\iota}$, $R_{n}(\bm{\beta}(\iota))$ is three times differentiable with respect to $\bm{\beta}(\iota)$ (see also Assumption \ref{assumption:estimation}). 
By ignoring small-order terms in Eq.~\eqref{eq:criteria:derivation:LapApp} and multiplying by $(-2)$, we obtain the more-extended criteria of the EBIC, as follows: 
\begin{eqnarray}
 - 2 \, M_{n}(\hat{\bm{\beta}}(\iota)) + \nu(\iota) \, \log n + 2 \, (1 - \gamma) \, \nu(\iota) \log P \:. \label{eq:criteria:EBICtype:def}
\end{eqnarray}
We refer to the criterion in Eq.~\eqref{eq:criteria:EBICtype:def} built upon statistical divergence excepting the KL divergence, such as the BHHJ divergence, as {\em E-DBBC} (extended DBBC). 

Noting that, although the remaining terms in Eq.~\eqref{eq:criteria:derivation:LapApp} are small-order with respect to $n$, the prior density $\pi_{\iota}$ and $\frac{\nu(\iota)}{2} \log ( 2 \, \pi )$ depend on $\nu(\iota)$, and the size of the matrix $\bm{T}_{n}(\hat{\bm{\beta}}(\iota))$ is $\nu(\iota)$ $\times$ $\nu(\iota)$. 
Since we consider models $\{ \mathcal{M}_{\iota} \}_{\iota}$ with $\nu(\iota)$ $\leq$ $S$ and $S$ $=$ $O(s)$ diverges to infinity, the $\nu(\iota)$ can diverge to infinity as $n$ goes to infinity. 
Therefore, we further propose another model evaluation criterion, without ignoring such terms, as follows: 
\begin{eqnarray}
 && - 2 \, R_{n}(\hat{\bm{\beta}}(\iota)) + \nu(\iota) \log n - \nu(\iota) \log \left( 2 \, \pi \right) \nonumber \\
 && \quad+\, \log \left| \bm{T}_{n}(\hat{\bm{\beta}}(\iota)) \right| + 2 \, (1 - \gamma) \, \nu(\iota) \log P \:, \label{eq:criteria:GEBICtype:def}
\end{eqnarray}
This criterion can be regarded as a generalization of a hybrid criterion of the EBIC and the GBIC (\cite{konishi2004bayesian}) along wide classes of divergence, thus we refer to the criterion in Eq.~\eqref{eq:criteria:GEBICtype:def} based on the KL divergence as {\em GEBIC}, and we name the criterion based on other divergence {\em GE-DBBC} (generalized E-DBBC). 
Since with the BIC, we choose a model (equivalent to determine a regularization parameter) whose value of Eq.~\eqref{eq:criteria:GEBICtype:def} is minimum in the candidate models. 
The characteristics of criteria, E-DBBC and GE-DBBC, can considerably depend on which term (divergence) we employ as the quasi-likelihood. 
For example, when we utilize a robust divergence like the BHHJ divergence, the criteria is expected to perform robustness in selection of the regularization parameter.

\subsection{Robustness of the criteria based on statistical divergence}
\label{sec:criteria:robustness}

In this subsection, we assess the robustness in variable selection against outliers of the three types of model evaluation criteria, DBBC, E-DBBC, and GE-DBBC, for some statistical divergence families, to achieve the aim (i) in Section \ref{sec:intro}. 
For each $i$ $=$ $1 \ldts n$, let $G_{i}$ be the ``true'' data-generating distribution of the response variable $Y_{i}$ with probability (density) function $g_{i}$, and let $\Xi_{i}$ be another distribution with probability (density) function $\xi_{i}$. 
Then, we define a mixture distribution $\Psi_{i}$ $=$ $(1 - r) \, G_{i}$ $+$ $r \, \Xi_{i}$ ($i$ $=$ $1 \ldts n$), for a (small) contamination rate $r$. 
Our goal is to model the ``true distribution'' $G$'s adequately, and we regard the ``another distribution'' $\Xi$'s as the source of outliers: we can see that, ``$r$ $=$ $0$'' corresponds to the case when there is no outlier in the data, while ``$r$ $>$ $0$'' means that a small proportion of the data is outlying. 
When $r$ $>$ $0$, it is expected that the behavior a robust analytical method will be similar to that when $r$ $=$ $0$. 
In parametric estimation, the influence functions and gross error sensitivity (\cite{huber1964robust, hampel1986robust}), that assess the damage caused by an outlier to an estimator, have been utilized in many fields. 

Recently, \cite{kurata2024robustness} introduced a measure of the sensitivity of a model evaluation criterion from the viewpoint of the difference of values of the criterion between the populations without outliers ($\bm{G}$ $=$ $( G_{1} \ldts G_{n} )$) and the contaminated populations by outlier-generation distributions ($\bm{\Psi}$ $=$ $( \Psi_{1} \ldts \Psi_{n} )$). 
Let $\mathcal{C}(\bm{G})$ be the value of a model evaluation criterion when the data generation distribution is $\bm{G}$ (i.e., when $r$ $=$ $0$). 
We can interpret that a criterion is sensitive against contamination of the data-generation distribution if (the absolute value of) the difference $\mathcal{I}(\bm{G} ,\, \bm{\Xi} ,\, r)$ $=$ $\mathcal{C}(\bm{\Psi})$ $-$ $\mathcal{C}(\bm{G})$ will be unboundedly large depending on the outliers. 
By contrast, we can regard a criterion as robust if $\mathcal{I}$ will be bounded for any outlier-generating distributions $\bm{\Xi}$ $=$ $( \Xi_{1} \ldts \Xi_{n} )$. 
To evaluate $\mathcal{I}$, we apply the second-order Taylor expansion around $r$ $=$ $0$. 
We here provide a theorem on the boundedness of the second-order approximation term. 

\begin{theorem}
\label{theorem:robustness}
Under (A-1)--(A-6) in Assumption \ref{assumption:estimation}, the DBBC, E-DBBC, GE-DBBC built upon the BHHJ divergence with $\alpha$ $>$ $0$ and JHHB divergence with $\alpha$ $>$ $0$ and $\varphi > 0$ satisfy that the second-order approximation term of $\mathcal{I}$ is bounded for arbitrary $\bm{\Xi}$. 
For the criteria built upon the C divergence, the boundedness holds if we further assume $\alpha$ $\geq$ $1$ and $\int \xi_{i}(y)^{2} \diff y$ $<$ $+\infty$ for each $i$ $=$ $1 \ldts n$. 
\end{theorem}

Theorem \ref{theorem:robustness} describes the robustness in model selection based on the BHHJ, JHHB, and C divergence families. 
This theorem also provides the conditions to achieve the aim (i) in Section \ref{sec:intro}, but the conditions differ depending on the class of divergence measures. 
The condition ``$\alpha$ $\geq$ $1$'' assigned for most of the divergence in C family is not desirable from the viewpoint of efficiency of the estimator, because using larger $\alpha$ in the C divergence family increases the asymptotic variance of the estimator and brings unstable estimation (refer also to, e.g., \cite{basu1998robust, maji2019robust}). 
Remark that, the BHHJ divergence is an exception; it requires only ``$\alpha$ $\geq$ $0$'' (see also \cite{kurata2024robustness}). 
Additionally, in continuous populations, we need to replace the true distribution $\bm{G}$ with some kind of estimated distribution such as kernel density when using the C divergence (see also \cite{maji2019robust}), whereas the BHHJ divergence does not require such procedure. 
Furthermore, the KL divergence and the JHHB divergence with $\varphi$ $=$ $0$ do not have the above-mentioned robustness, and the JHHB divergence with $\varphi$ $>$ $0$ and $\varphi$ $\neq$ $1$ does not have exact unbiasedness in estimation (\cite{jones2001comparison}). 
In light of the above, Theorem \ref{theorem:robustness} shows the superiority of the BHHJ divergence in the JHHB and C divergence families regarding to the robust selection of the regularization parameter and explanatory variables. 
We will prove this theorem in Section \ref{secappend:proofs}: this theorem can be proved going along with the proofs of Theorems 1--3 in \cite{kurata2024robustness}. 
It should be noted that, Theorem \ref{theorem:robustness} does not restrict the asymptotic behavior (divergence speed) of the number of candidate explanatory variables ($P$) and the necessary variables ($s$), therefore, this theorem guarantees the robustness of the three types of criteria, DBBC, E-DBBC, and GE-DBBC in not only the case of fixed dimensionality but also high-dimensional settings.

\subsection{Model selection consistency of the criteria based on statistical divergence in the high-dimensional regression}
\label{sec:criteria:consistency}

The selection consistency is a meaningful property for the purpose of detecting necessary variables in the various candidates, and it can be compatible with robustness against outliers. 
Selecting a too large value of the regularization parameter brings an under-specified model that misses some of necessary explanatory variables, and a too small value makes an over-specified model that employs even unnecessary variables. 
In the case of fixed dimensionality, \cite{kurata2020consistency, kurata2024robustness} proposed robust model evaluation criteria with selection consistency, based on divergence measures. 
In this subsection, we provide a theorem supporting the selection consistency of the proposed criteria even in the high-dimensional setting, to achieve the aim (ii) in Section \ref{sec:intro}. 
Regarding this, we give some conditions in order to prove asymptotic properties in Subsection \ref{secappend:conditions:Cz}. 

\begin{theorem}
\label{theorem:consistency}
Under all conditions in Assumption \ref{assumption:estimation} and Assumption \ref{assumption:consistency}, and supposing $S$ $=$ $o( n^{\min \{ l / 2 ,\, (1 - l) / 2 \}} )$, the E-DBBC and GE-DBBC have the selection consistency, i.e., for any under- and over-specified model $\mathcal{M}_{\iota}$ with $\nu(\iota)$ $\leq$ $S$, the probabilities 
$\prob \{ \EDBBC(\mathcal{M}_{\iota})$ $-$ $\EDBBC(\mathcal{M}_{\iota_{\star}})$ $>$ $0 \}$ and $\prob \{ \GEDBBC(\mathcal{M}_{\iota})$ $-$ $\GEDBBC(\mathcal{M}_{\iota_{\star}})$ $>$ $0 \}$
converge to one as $n$ goes to $+\infty$. 
\end{theorem}

Theorem \ref{theorem:consistency} guarantees the selection consistency of the E-DBBC and GE-DBBC in high-dimensional settings, but does not support that of the DBBC including the BIC. 
Actually, DBBC-type criteria tend to select over-specified models probabilistically in high-dimensional settings, even if the sample size $n$ is sufficiently large. 
The proof will be provided in Section \ref{secappend:proofs}.

\subsection{On the form of the weight function and the asymptotic order of the regularization parameter}
\label{sec:criteria:goodexample}

\begin{figure}[ht]
\begin{center}
\includegraphics[width=0.55\linewidth]{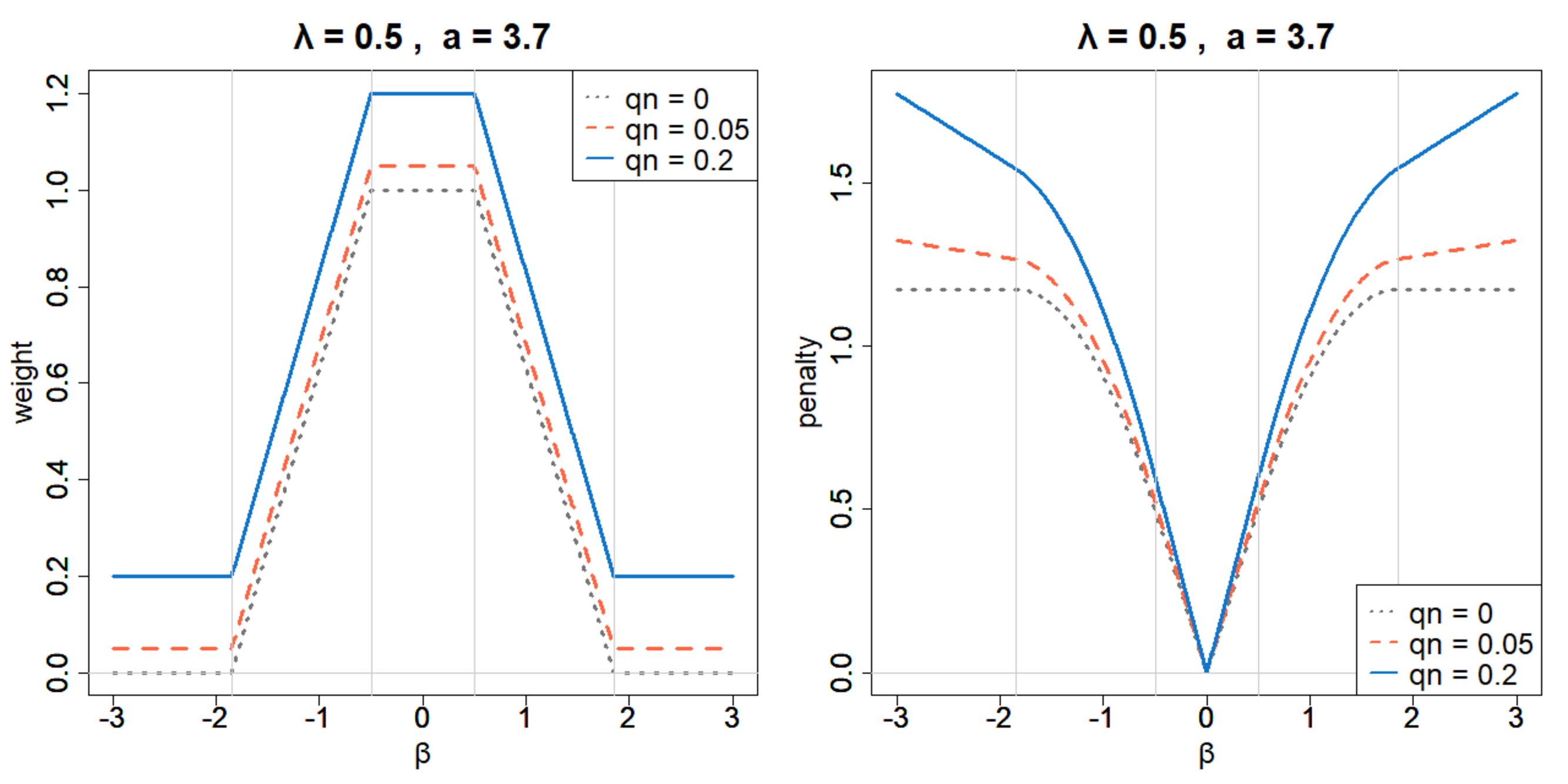}
\caption{The weight function and corresponding penalty function for $q_{n}$ $=$ $0$ (grey dotted line: corresponding the SCAD penalty), $0.05$ (red dashed line), $0.20$ (blue solid line), with $\lambda$ $=$ $0.5$ and $a$ $=$ $3.7$. Grey vertical lines indicate $\beta$ $=$ $\pm \, \lambda$ and $\pm \, a \, \lambda$, the change points of the penalty (see also Eqs.~\eqref{eq:weights:SCADdiff} and \eqref{eq:weights:SCADQ}). }
\label{fig:SCADQ}
\end{center}
\end{figure}

When analyzing data by the adaptive LASSO, the choice of the weights $w_{j}$ $=$ $h(|\hat{\beta}^{\iniest}_{j}|)$ ($j$ $=$ $1 \ldts P$) is pivotal as well as the determination of the regularization parameter. 
As the weight function $h(|\hat{\beta}^{\iniest}_{j}|)$, reciprocal function $1 \,/\, |\hat{\beta}^{\iniest}_{j}|$ or its exponentiation have been often used (e.g., \cite{chatterjee2013rates, huang2008adaptive, zou2006adaptive}). 
Moreover, the first-derivative of the SCAD penalty (\cite{fan2001variable}), 
\begin{eqnarray}
 \tau(|\beta|) \:=\: \left\{ \begin{array}{ll}
  1 & \left( |\beta| \,\leq\, \lambda \right) \\
  \frac{a \, \lambda - |\beta|}{(a - 1) \, \lambda} & \left( \lambda \,\leq\, |\beta| \,\leq\, a \, \lambda \right) \\
  0 & \left( a \, \lambda \,\leq\, |\beta| \right) \\
 \end{array} \right. \:, \label{eq:weights:SCADdiff}
\end{eqnarray}
has also been utilized as the weight function (e.g., \cite{fan2014adaptive, ghosh2024robust}), where $a$ $>$ $2$ is a some constant ($a$ $=$ $3.7$ is commonly used: see \cite{fan2001variable}). 
Since every element in $\bm{\beta}_{\star}^{(1)}$ is non-zero and $\bm{\beta}_{\star}^{(2)}$ $=$ $\bm{0}_{P-s}$, we can expect that the values of $\tau(|\hat{\beta}^{\iniest}_{j}|)$ will be equal or close to $0$ for any $j$ $\in$ $\mathcal{J}^{(1)}$ and will be equal or close to $1$ for $j$ $\in$ $\mathcal{J}^{(2)}$. 
Noting that, some weight functions including the reciprocal-type functions do not fulfill the sufficient conditions for establishing the oracle property (see also Subsection \ref{secappend:conditions:Az}). 
On the other hand, the SCAD-based weight function in Eq.~\eqref{eq:weights:SCADdiff} is desirable from the viewpoint of the oracle property (see also \cite{fan2014adaptive, ghosh2024robust}), nevertheless, there is a problem when we apply it to the model evaluation criteria. 
The density function of prior distribution, $\pi_{\iota}$ (Laplace distribution: see Eq.~\eqref{eq:criteria:GALASSOprior}), will be equal to $0$ if $w_{j}$ $=$ $0$ holds for any one of $j$ $=$ $1 \ldts P$. 
In such a case, we can not obtain $\log \pi_{\iota}$, and thus we can not derive the GE-DBBC defined in Eq.~\eqref{eq:criteria:GEBICtype:def}. 
Therefore, we here employ another weight function, as follows: 
\begin{eqnarray}
 w_{j} \:=\: h(|\hat{\beta}^{\iniest}_{j}|) \:=\: q_{n} + \tau(|\hat{\beta}^{\iniest}_{j}|) \quad \left( j \,=\, 1 \ldts P \right) \:, \label{eq:weights:SCADQ}
\end{eqnarray}
where $q_{n}$ is some positive value fulfilling $q_{n}$ $\to$ $0$ as $n$ $\to$ $+\infty$. 
Clearly, this weight function can be regarded as a generalization of the SCAD. 
The SCAD penalty behaves like the $L^{1}$-penalty in a neighborhood of the origin, has the form of a quadratic function for values slightly away from the origin, and takes a constant value (i.e., behaves like the $L^{0}$-penalty) for values further away from the origin (see also \cite{fan2001variable, fan2020statistical}). 
We can see that, the adaptive LASSO via the weight function defined in Eq.~\eqref{eq:weights:SCADQ} is equivalent to the first-order approximation of the penalty function that behaves like the $L^{1}$-penalty in a neighborhood of the origin, has the form of a quadratic function for values slightly away from the origin, and behaves like the small-slope $L^{1}$-penalty for values further away from the origin (Fig.~\ref{fig:SCADQ}). 

For achieving the two main aims of this paper, (i) and (ii) simultaneously, the convergence speed of $q_{n}$ in Eq.~\eqref{eq:weights:SCADQ} will be restricted. 
We now provide the sufficient conditions for fulfilling Assumptions \ref{assumption:estimation} and \ref{assumption:consistency}, when employing $h$ defined in Eq.~\eqref{eq:weights:SCADQ} as the weight function. 

\begin{corollary}
\label{corollary:SCADQandBHHJ}
Suppose (A-1)--(A-3) in Assumption \ref{assumption:estimation} and (C-1)--(C-3) in Assumption \ref{assumption:consistency}. 
Then, in the case of the normal linear regression model, the regularization method employing the weight function in Eq.~\eqref{eq:weights:SCADQ} with $q_{n}$ $\leq$ $\frac{a_{4}}{\sqrt{s \, \log P}}$ for a positive constant $a_{4}$ (given in Assumption \ref{assumption:estimation}) and the model selection via the E-DBBC and GE-DBBC based on the BHHJ divergence achieve the same properties as those described in Proposition \ref{proposition:estimation}, Theorem \ref{theorem:robustness}, and Theorem \ref{theorem:consistency}. 
\end{corollary}

Corollary \ref{corollary:SCADQandBHHJ} shows that the functional form of Eq.~\eqref{eq:weights:SCADQ} is one of desirable weights that achieve the oracle property and enable us to derive model evaluation criteria naturally. 
Remark that, asymptotic properties such as the oracle property do not hold if we employ too large $q_{n}$ ($\not \to 0$). 
As the value of $q_{n}$, we can use some sufficiently small positive value in practice, for example, $q_{n}$ $=$ $n^{-2}$. 
We can see that, assumptions about the regularization parameter $\lambda$ $=$ $\lambda_{n}$ when it has the order $\lambda$ $=$ $O( \sqrt{\frac{s \, \log P}{n}} )$ (see also \cite{ghosh2024robust}). 
However, since the oracle property holds for any values of $\lambda$ that are proportional to $\sqrt{\frac{s \, \log P}{n}}$, we need to determine a single value in order to obtain the optimal combination of the explanatory variables, as previously stated in this paper.

\subsection{On the differences among three types of criteria}
\label{sec:criteria:diffICs}

We here discuss the difference of the three types of criteria, the DBBC, E-DBBC, and GE-DBBC. 
Although both of the E-DBBC and GE-DBBC have the selection consistency as mentioned in Theorem \ref{theorem:consistency}, their additional terms are different to no small extent. 
In this subsection, we denote the value of a criterion with respect to model $\mathcal{M}_{\iota}$ by 
\begin{eqnarray}
 \Criterion(\mathcal{M}_{\iota}) \:=\: - 2 \, M_{n}(\hat{\bm{\beta}}(\iota)) \,+\, A^{\Criterion}_{n} \:, \nonumber
\end{eqnarray}
and refer to $A^{\Criterion}_{n}$ as the additional term of the criterion; we can see that the additional terms corresponding to the DBBC, E-DBBC, and GE-DBBC, are 
\begin{eqnarray}
 A^{\DBBC}_{n} &\bseq& \nu(\iota) \log n \:, \nonumber \\
 A^{\EDBBC}_{n} &\bseq& \nu(\iota) \log n + 2 \, (1 - \gamma) \, \nu(\iota) \log P \:, \nonumber \\
 A^{\GEDBBC}_{n} &\bseq& - 2 \, \log \pi_{\iota}(\hat{\bm{\beta}}(\iota)) + \nu(\iota) \log n - \nu(\iota) \log \left( 2 \, \pi \right) \nonumber \\
 && \;+\, \log \left| \bm{T}_{n}(\hat{\bm{\beta}}(\iota)) \right| + 2 \, (1 - \gamma) \, \nu(\iota) \log P \:, \label{eq:additerm:GEDBBC}
\end{eqnarray}
respectively, where $\pi_{\iota}$ is the probability density function of the prior distribution (in our case, it corresponds to the Laplace distribution: see also Eq.~\eqref{eq:criteria:GALASSOprior}). 
Now, we show the asymptotic behavior of each term in Eq.~\eqref{eq:additerm:GEDBBC}. 
We can see that $A^{\GEDBBC}_{n}$ can be reduced as follows (see also Eq.~\eqref{eq:criteria:GALASSOprior}): 
\begin{eqnarray}
 A^{\GEDBBC}_{n} &\bseq& 2 \, n \, \lambda \sum_{j \in \mathcal{J}(\iota)^{(1)}} w_{j} \, \left| \hat{\beta_{j}}(\iota) \right| - 2  \sum_{j \in \mathcal{J}(\iota)^{(1)}} \log w_{j} + \log \left| \bm{T}_{n}(\hat{\bm{\beta}}(\iota)) \right| \nonumber \\
 && \;+\, \nu(\iota) \left\{ - 2 \, \log \frac{n \, \lambda}{2} + \log n - \log \left( 2 \, \pi \right) + 2 \, (1 - \gamma) \, \log P \right\} \:. \label{eq:additerm:GEDBBC:reduced1}
\end{eqnarray}
If we employ the function given in Eq.~\eqref{eq:weights:SCADQ} as the weight $w_{j}$ and assume $q_{n}$ $=$ $n^{-\zeta}$ ($\zeta$ $>$ $3/2$), the first term in Eq.~\eqref{eq:additerm:GEDBBC:reduced1} will go to $0$ as $n$ $\to$ $+\infty$ for sufficiently large $n$, because every true value of the regression coefficient is in constant-order (or equal to $0$), moreover, we can confirm that $- 2 \, \sum_{j \in \mathcal{J}(\iota)^{(1)}} \log w_{j}$ is asymptotically equivalent to $2 \, \nu(\iota) \, \zeta \, \log n$. 
Therefore, under our settings, we can reduce $A^{\GEDBBC}_{n}$ as 
\begin{eqnarray}
 \nu(\iota) \left\{ - 2 \, \log \lambda + (2 \, \zeta - 1) \, \log n + 2 \, (1 - \gamma) \, \log P \right\} + \log \left| \bm{T}_{n}(\hat{\bm{\beta}}(\iota)) \right| + o_{P}(1) \:. \label{eq:additerm:GEDBBC:reduced2}
\end{eqnarray}
Additionally, by (C-4) in Assumption \ref{assumption:consistency}, $\log | \bm{T}_{n}(\hat{\bm{\beta}}(\iota)) |$ is asymptotically proportional to (or less than) $\nu(\iota)$. 
If we further specify that the regularization parameter $\lambda$ $=$ $\lambda_{n}$ is asymptotically proportional to $\sqrt{\frac{\nu(\iota) \, \log P}{n}}$ (see also Subsection \ref{sec:criteria:goodexample}), then Eq.~\eqref{eq:additerm:GEDBBC:reduced2} will be evaluated as 
\begin{eqnarray}
 \nu(\iota) \left\{ \left( 2 \, \zeta - l \right) \, \log n + 2 \, (1 - \gamma) \, \log P - \log \nu(\iota) + O_{P}(1) \right\} + o_{P}(1) \:, \label{eq:additerm:GEDBBC:reduced3}
\end{eqnarray}
since $\log \log P$ $=$ $l \, \log n$ ($l \in (0 ,\, 1)$). 
From the form of Eqs.~\eqref{eq:additerm:GEDBBC:reduced2} and \eqref{eq:additerm:GEDBBC:reduced3}, 
\begin{eqnarray}
 A^{\GEDBBC}_{n} \,>\, A^{\DBBC}_{n} \:,\quad A^{\GEDBBC}_{n} \,>\, A^{\EDBBC}_{n} \nonumber
\end{eqnarray}
hold for sufficiently large $n$. 
From these, we can see that more precise approximation for the (quasi) posterior probability leads more severe evaluation for the over-specified models. 
In other words, the GE-DBBC tends not to select the over-specified models relatively to the DBBC and E-DBBC when the sample size $n$ is large.

\section{Numerical examples}
\label{sec:numerical}

In the previous section, we showed that three types of the model evaluation criteria, the DBBC, E-DBBC, and GE-DBBC, achieve the robustness in model selection when we employ appropriate divergence measures, and that the E-DBBC and GE-DBBC also have the selection consistency even in the high-dimensional setting under some regularity conditions. 
In this section, we introduce some numerical examples to verify the properties of the proposed criteria. 
Considering the superiority in robustness described in Theorem \ref{theorem:robustness}, we employed the BHHJ divergence (Eq.~\eqref{eq:BHHJdiv}) to derive the model evaluation criteria. 
As the weights of the adaptive LASSO, we utilized the function defined in Eq.~\eqref{eq:weights:SCADQ} with $q_{n}$ $=$ $n^{-2}$. 
Throughout this section, we use the criterion that has the additional term given in Eq.~\eqref{eq:additerm:GEDBBC:reduced2} omitting the part of $o_{P}(1)$, as the GE-DBBC. 

\subsection{Simulations}
\label{sec:numerical:simulation}

We conducted simulation studies to verify the robustness against outliers and the selection consistency. 
We generated the explanatory variables $\bm{x}_{1} \ldts \bm{x}_{n}$ independently from the normal distribution $\normaldist ( \bm{0}_{P} ,\, \Imat_{P \times P} )$: the value of the design matrix $\bm{X}$ $=$ $( \bm{x}_{1} \ldts \bm{x}_{n} )^{\top}$ was fixed through the simulations. 
We defined the true value of the regression coefficient $\bm{\beta}_{\star}$ as $( \bm{1}_{5}^{\top} ,\, \bm{0}_{P-5}^{\top} )^{\top}$, i.e., $s$ $=$ $5$, and we generated $n$ centered response variables according to the normal linear regression model (Eq.~\eqref{eq:NormalLinearReg}) with $\sigma^{2}$ $=$ $1$. 
We set four pairs of the sample sizes $n$ and the number of candidate explanatory variables $P$, $(n ,\, P)$ $=$ $(5000 ,\, 20)$, $(5000 ,\, 200)$, $(500 ,\, 200)$, and $(200 ,\, 1000)$. 
Then, to measure the selection robustness, we replaced values of $Y$'s with ten times of their original values with probability $r$. 
As the value of $r$, we used $0$ (non-contaminated setting), $0.01$, $0.05$, and $0.10$. 
For comparison, we also conducted variable selection via the BIC and EBIC using the original LASSO (uniform weight): we refer as them to ``unif-BIC'' and ``unif-EBIC'', respectively. 

\begin{figure}[ht]
\begin{center}
\includegraphics[width=0.95\linewidth]{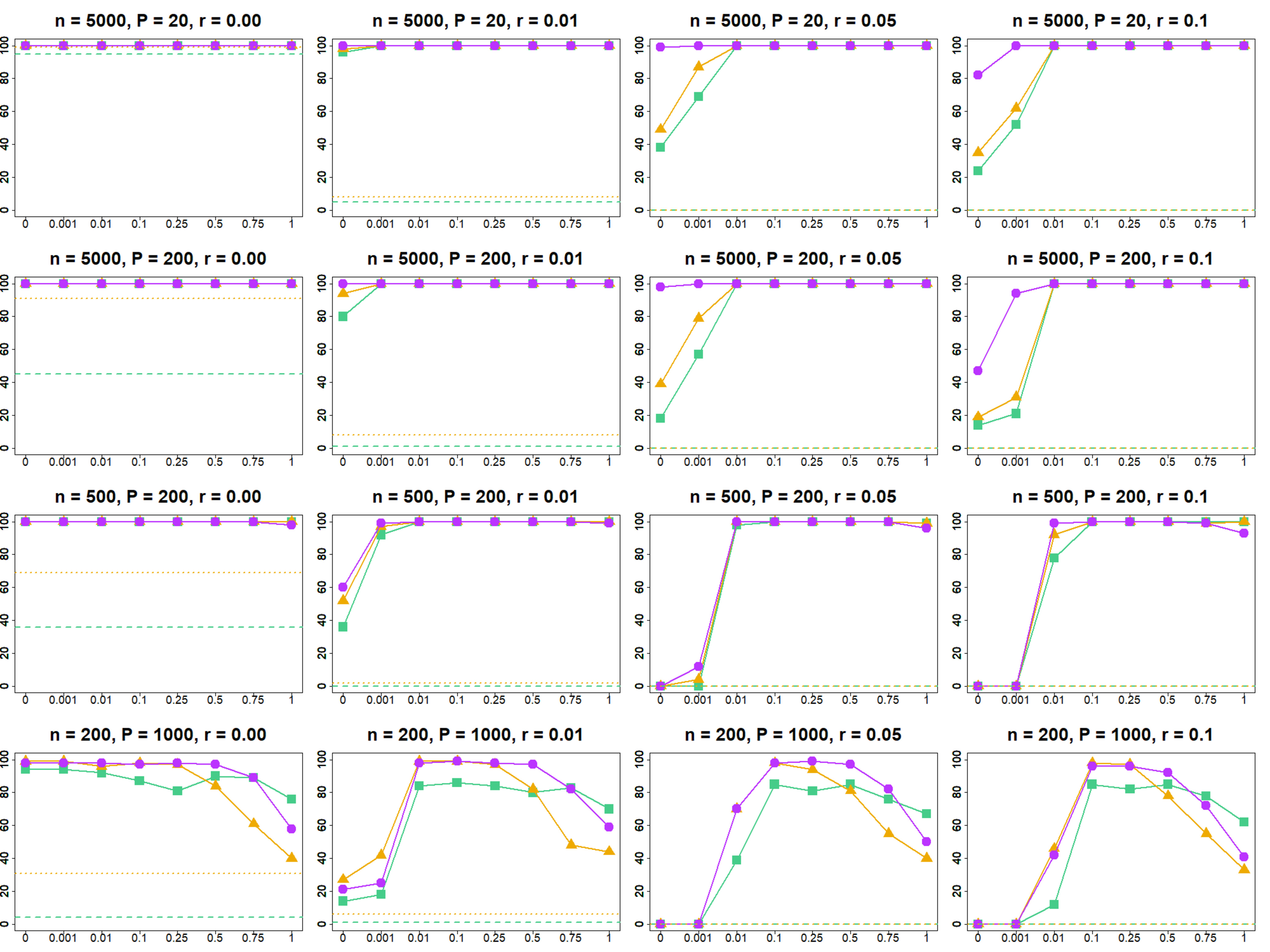}
\caption{The accuracy rates (\%) of the DBBC (green line graphs with square symbols), E-DBBC (orange, triangle symbols), and GE-DBBC (purple, circle symbols) with $\alpha$ $=$ $0$, $0.001$, $0.01$, $0.1$, $0.25$, $0.5$, $0.75$, and $1$, for each sample size $n$, the number of candidate explanatory variables $P$, and contamination rate $r$. Dashed horizontal line indicates the accuracy rates of the unif-BIC (green) and unif-EBIC (orange). }
\label{fig:simulation:main}
\end{center}
\end{figure}

Fig.~\ref{fig:simulation:main} illustrates the accuracy rates of $100$ trials of the unif-BIC, unif-EBIC and three types of criteria (DBBC, E-DBBC, and GE-DBBC) for some values of $\alpha$ ($0$, $0.001$, $0.01$, $0.1$, $0.25$, $0.5$, $0.75$, and $1$). 
Additionally, the detailed selection rates for each $(n ,\, P)$ are shown in Section \ref{secappend:tables}. 
From the figure and tables, all criteria performed well in the non-contaminated setting ($r$ $=$ $0$) with $(n ,\, P)$ $=$ $(5000 ,\, 20)$, however, in other settings, the accuracy greatly varies depending on the type of criterion and the value of the tuning parameter $\alpha$. 
We can see that the unif-BIC and unif-EBIC did not performed well when $P$ was not small or there existed some outliers in the data. 
As the unif-BIC and unif-EBIC are based on the KL divergence, they do not have the robustness in estimation and model selection. 
Additionally, the LASSO-type estimator employing the uniform weight does not fulfill (A-4) and (A-5) in Assumption \ref{assumption:estimation}, hence Theorem \ref{theorem:consistency} does not hold for the unif-BIC and unif-EBIC (in the first place, the uniform weight does not achieve the oracle property: see also, e.g., \cite{fan2001variable, zou2006adaptive}). 
Thus, we can not support that these two criteria have either the robustness in model selection or selection consistency in high-dimensional settings. 

The comparison among $\alpha$ of the three types of criteria (DBBC, E-DBBC, and GE-DBBC) reflects the effects of robustness of the BHHJ divergence. 
In the non-contaminated settings, every criterion performed well, yet we can see that criteria with small $\alpha$ failed to select the true model in contaminated settings. 
Particularly, most criteria with $\alpha$ $=$ $0$ (based on the KL divergence) and $\alpha$ $=$ $0.001$ selected over-specified models at high rates in heavily contaminated settings (see also the columns of ``$r$ $=$ $0.05$'' and ``$r$ $=$ $0.1$'' in Tables \ref{tab:simulation:main:n5000P20}--\ref{tab:simulation:main:n200P1000} in \ref{secappend:tables}). 
By contrast, we can confirm that the criteria with moderately large $\alpha$ $>$ $0$ performed well regardless of whether there existed outliers in the observations or not. 
This results were consistent with Theorem \ref{theorem:robustness} describing that the model evaluation criteria based on appropriate divergence measures including the BHHJ divergence are robust against the contamination of outliers. 
For $(n ,\, P)$ $=$ $(200 ,\, 1000)$, criteria with $\alpha$ $>$ $0.5$ did not perform well in comparison with the cases for large $n$. 
In fact, larger $\alpha$ of the C divergence family (including the BHHJ divergence) increases the asymptotic variance of the estimator (\cite{basu1998robust, maji2019robust, vonta2012properties}), hence an excessively large $\alpha$ is not desirable from the viewpoint of parametric estimation. 
In previous studies, estimation and model selection with too large $\alpha$ also tend to be unstable due to the lack of efficiency, especially when the sample size is not sufficiently large (e.g., \cite{basu1998robust, ghosh2013robust, kurata2024robustness}). 
On detailed discussion about the optimal value or range of the tuning parameter, refer also to, e.g., \cite{basak2021optimal, ghosh2015robust, warwick2005choosing}. 

Results for $(n ,\, P)$ $=$ $(200 ,\, 1000)$ suggests that the DBBC (including the BIC) has a drawback that can not cope well with the case where the number of candidate explanatory variables is large and the sample size is not sufficiently large. 
These results agreed with Theorem \ref{theorem:consistency} and the description in Subsection \ref{sec:criteria:diffICs}. 
The DBBC has asymptotically smaller additional term than the E-DBBC and GE-DBBC (see Eqs.~\eqref{eq:additerm:GEDBBC} and \eqref{eq:additerm:GEDBBC:reduced2}), and thus selection via the DBBC tends fail to remove some unnecessary variables in high-dimensional settings. 
The E-DBBC and GE-DBBC recorded similar results, yet the GE-DBBC performed better in most of contaminated settings. 
In particular, the GE-DBBC tended to prevent to select over-specified models relatively to the DBBC and E-DBBC, for heavily contaminated cases. 
We can consider that more precise approximation for the quasi-posterior probability and the asymptotically large additional term of the GE-DBBC brought the superior performance.

\subsection{Real data analysis}
\label{sec:numerical:realdata}

\begin{figure}[ht]
\begin{center}
\includegraphics[width=0.75\linewidth]{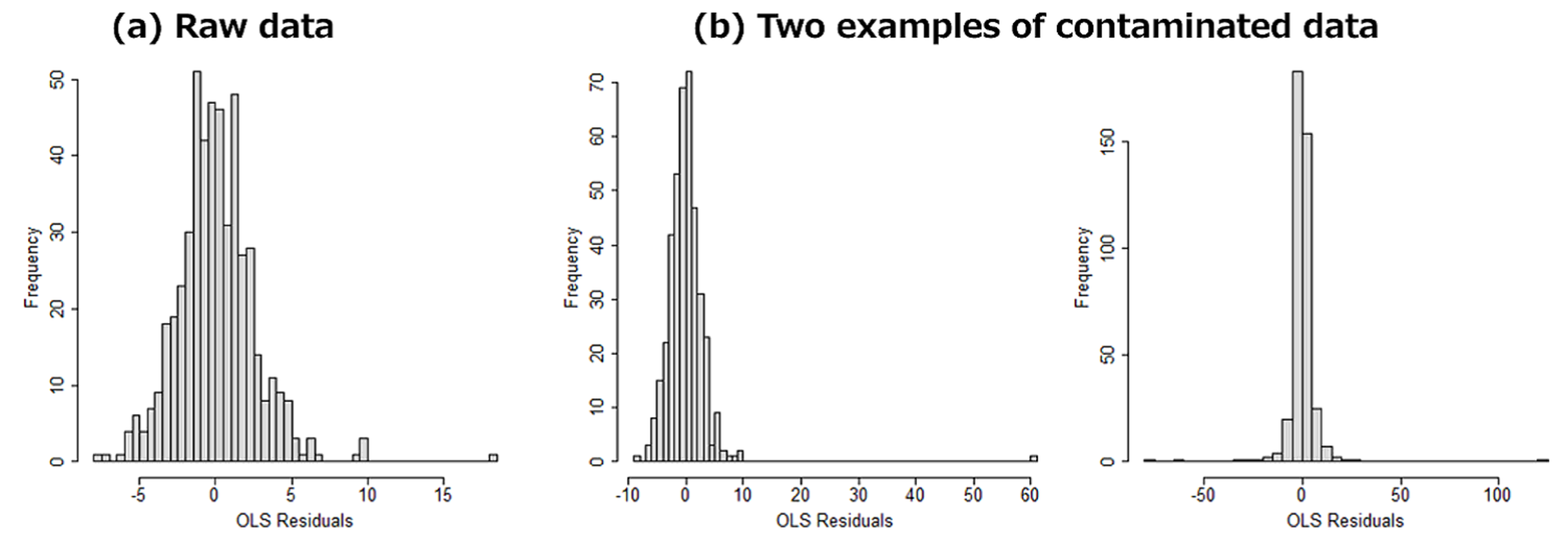}
\caption{Histograms of the residuals for the OLS estimates using all explanatory variables, for (a) raw data and (b) contaminated training data. }
\label{fig:real:hist}
\end{center}
\end{figure}

In this subsection, we show an application of the proposed model evaluation criteria for the variable selection problem using the Boston housing data set. 
The dataset consists of $506$ observations and $14$ variables (on the detail of this dataset, see \cite{harrison1978hedonic}). 
We employed the median value of owner-occupied homes as the response variable (centered), and other variables (standardized) and their interaction terms as explanatory variables: we conducted variable selection to detect the set necessary explanatory variables from the $P$ $=$ $91$ candidates. 
We randomly split the dataset into a training data and test data at a ratio of $4$:$1$ (the sample size of the training data was $n$ $=$ $405$). 
As the value of the $\sigma^{2}$ (variance of the error terms) for every method, we used the square of the adjusted standard deviation of the residuals with respect to the LASSO estimates, to ensure fairness among the selection methods. 
Fig.~\ref{fig:real:hist}(a) is the histogram of the residuals for the OLS estimates when employing all $P$ explanatory variables. 
We can see that the residuals roughly distributed symmetrically and suspect that there can exist some outliers in the observations. 

Next, in order to verify the robustness in selection, we randomly replaced $k$ values of the response variables in the training data with $m$ times of its original value. 
Fig.~\ref{fig:real:hist}(b) shows the examples of histogram of the residuals for the OLS estimates (using all explanatory variables), for contaminated data when ($k$, $m$) $=$ ($1$, $10$). 
We can see more outlying residuals that exert negative effect on estimation and variable selection, in comparison with the raw data. 
We can consider that outliers distorted the (non-robust) OLS estimation, and brought the extremely large residuals. 
In this experiment, we employed $\alpha$ $=$ $0.1$ as the value of the tuning parameter of the BHHJ divergence, taking the stable performance in the simulation into account (Subsection \ref{sec:numerical:simulation}). 
To measure the robustness of each model criterion against the contamination of outliers, we define ``CR'', the concordance rate with the selected result of the raw data. 
If a criterion selects similar explanatory variables for both of the raw and contaminated data, the value of CR will be close to $1$ (see Table \ref{tab:ConcordanceRate}). 

\begin{table}[ht]
    \caption{The concordance rate (CR) is calculated as (A + D) / (A + B + C + D). }
    \label{tab:ConcordanceRate}
        \begin{tabular}{|cc|cc|} \hline
             &  & \multicolumn{2}{c|}{Contaminated data} \\
             &  & Necessary & Unnecessary \\ \hline
            \multirow{2}{*}{Raw data} & Necessary & \textcolor[rgb]{0.8,0,0}{A} & C \\
             & Unnecessary & B & \textcolor[rgb]{0.8,0,0}{D} \\ \hline
            \end{tabular}
\end{table}

Tables \ref{tab:realsummary:10bai1ko}, \ref{tab:realsummary:10bai5ko}, and \ref{tab:realsummary:20bai1ko} illustrate the summaries of results (the RMSE of the residuals of test data, the number of the selected explanatory variables, and the CR) of $100$ trials of the three types of criteria (DBBC, E-DBBC, and GE-DBBC) utilizing the BHHJ divergence with $\alpha$ $=$ $0$ and $0.1$, for ($k$, $m$) $=$ ($1$, $10$), ($5$, $10$), and ($1$, $20$), respectively. 
For the raw data, the DBBC with $\alpha$ $=$ $0$ (corresponding to BIC) and E-DBBC with $\alpha$ $=$ $0$ (EBIC), and E-DBBC with $\alpha$ $=$ $0.1$ and GE-DBBC with $\alpha$ $=$ $0.1$ selected the same combinations of explanatory variables, respectively. 
Regarding the average, the RMSEs of the residuals of test data for $\alpha$ $=$ $0.1$ were not much different from for $\alpha$ $=$ $0$, but the standard deviations for $\alpha$ $=$ $0.1$ were remarkably smaller than for $\alpha$ $=$ $0$ when there existed multiple outliers (Table \ref{tab:realsummary:10bai5ko}) and there existed a more extreme outlier (Table \ref{tab:realsummary:20bai1ko}). 
We can consider that, non-robust criteria tended to have extremely large residuals due to outliers (see also histograms in Fig.~\ref{fig:real:hist}), as a result, the variances of the RMSE became large. 
Moreover, we can see that the values of CR of the criteria for $\alpha$ $=$ $0$ were smaller than those of the E-DBBC and GE-DBBC for $\alpha$ $=$ $0.1$: these reflect that non-robust criteria treated outliers equally with other observations, and thus they failed to perform consistent variable selection robustly. 
For both of raw data and contaminated data, the DBBC generally selected more candidate variables than the E-DBBC and GE-DBBC. 
This trend was similar to the previous simulations (see Subsection \ref{sec:numerical:simulation}), and agrees with the theoretical descriptions in this paper (Theorem \ref{theorem:consistency} and Subsection \ref{sec:criteria:diffICs}). 
Although the DBBC recorded smaller RMSE than the E-DBBC and GE-DBBC, there is suspicion that the DBBC selected not only truly necessary variables but also some unnecessary ones. 
High averages and small standard deviations of the CR for the E-DBBC and GE-DBBC with $\alpha$ $=$ $0.1$ show the robust and stable variable selection of the proposed method. 

\begin{table}[ht]
    \caption{Averages (Ave) and standard deviations (SD) of the root mean squared error of the residuals of the test data (RMSE), the number of the selected explanatory variables ($\nu$), and the concordance rate (CR), for the contaminated data with ($k$, $m$) $=$ ($1$, $10$). The column of ``Raw'' shows the results for the raw data. }
    \label{tab:realsummary:10bai1ko}
        \begin{tabular}{|l|rr|rrr|rr|} \hline
             & \multicolumn{2}{c|}{RMSE} & \multicolumn{3}{c|}{$\nu$} & \multicolumn{2}{c|}{CR} \\ \hline 
            Criterion ($\alpha$) & \multicolumn{1}{c}{Ave} & \multicolumn{1}{c|}{(SD)} & \multicolumn{1}{c}{Raw} & \multicolumn{1}{c}{Ave} & \multicolumn{1}{c|}{(SD)} & \multicolumn{1}{c}{Ave} & \multicolumn{1}{c|}{(SD)} \\ \hline
            DBBC ($0$)      & $4.67$ & ($1.50$) & $50$ & $40.57$ & ($10.61$) & $0.76$ & ($0.10$) \\
            E-DBBC ($0$)    & $4.75$ & ($1.35$) & $50$ & $31.04$ & ($ 8.28$) & $0.71$ & ($0.09$) \\
            GE-DBBC ($0$)   & $4.92$ & ($1.26$) & $38$ & $23.98$ & ($ 6.91$) & $0.78$ & ($0.07$) \\
            \hline          
            DBBC ($0.1$)    & $4.48$ & ($0.85$) & $45$ & $29.90$ & ($ 9.98$) & $0.78$ & ($0.08$) \\
            E-DBBC ($0.1$)  & $4.89$ & ($0.99$) & $18$ & $19.71$ & ($ 7.00$) & $0.91$ & ($0.05$) \\
            GE-DBBC ($0.1$) & $5.34$ & ($1.16$) & $18$ & $15.24$ & ($ 5.48$) & $0.92$ & ($0.03$) \\
            \hline          
        \end{tabular}
\end{table}

\begin{table}[ht]
    \caption{Results for ($k$, $m$) $=$ ($5$, $10$). }
    \label{tab:realsummary:10bai5ko}
        \begin{tabular}{|l|rr|rrr|rr|} \hline
             & \multicolumn{2}{c|}{RMSE} & \multicolumn{3}{c|}{$\nu$} & \multicolumn{2}{c|}{CR} \\ \hline 
            Criterion ($\alpha$) & \multicolumn{1}{c}{Ave} & \multicolumn{1}{c|}{(SD)} & \multicolumn{1}{c}{Raw} & \multicolumn{1}{c}{Ave} & \multicolumn{1}{c|}{(SD)} & \multicolumn{1}{c}{Ave} & \multicolumn{1}{c|}{(SD)} \\ \hline
            DBBC ($0$)      & $7.05$ & ($3.71$) & $50$ & $26.52$ & ($8.70$) & $0.60$ & ($0.06$) \\
            E-DBBC ($0$)    & $7.05$ & ($2.93$) & $50$ & $17.93$ & ($9.85$) & $0.57$ & ($0.06$) \\
            GE-DBBC ($0$)   & $7.13$ & ($2.69$) & $38$ & $12.17$ & ($8.16$) & $0.66$ & ($0.05$) \\
            \hline          
            DBBC ($0.1$)    & $5.47$ & ($1.27$) & $45$ & $15.36$ & ($7.05$) & $0.66$ & ($0.06$) \\
            E-DBBC ($0.1$)  & $6.51$ & ($1.49$) & $18$ & $ 9.58$ & ($6.39$) & $0.88$ & ($0.06$) \\
            GE-DBBC ($0.1$) & $7.06$ & ($1.42$) & $18$ & $ 6.97$ & ($4.81$) & $0.86$ & ($0.04$) \\
            \hline          
        \end{tabular}
\end{table}

\begin{table}[ht]
    \caption{Results for ($k$, $m$) $=$ ($1$, $20$). }
    \label{tab:realsummary:20bai1ko}
        \begin{tabular}{|l|rr|rrr|rr|} \hline
             & \multicolumn{2}{c|}{RMSE} & \multicolumn{3}{c|}{$\nu$} & \multicolumn{2}{c|}{CR} \\ \hline 
            Criterion ($\alpha$) & \multicolumn{1}{c}{Ave} & \multicolumn{1}{c|}{(SD)} & \multicolumn{1}{c}{Raw} & \multicolumn{1}{c}{Ave} & \multicolumn{1}{c|}{(SD)} & \multicolumn{1}{c}{Ave} & \multicolumn{1}{c|}{(SD)} \\ \hline
            DBBC ($0$)      & $5.80$ & ($3.26$) & $50$ & $34.11$ & ($12.30$) & $0.69$ & ($0.12$) \\
            E-DBBC ($0$)    & $5.86$ & ($2.65$) & $50$ & $24.50$ & ($11.58$) & $0.64$ & ($0.10$) \\
            GE-DBBC ($0$)   & $6.13$ & ($2.54$) & $38$ & $18.19$ & ($ 9.76$) & $0.73$ & ($0.09$) \\
            \hline          
            DBBC ($0.1$)    & $5.08$ & ($1.36$) & $45$ & $22.90$ & ($11.90$) & $0.72$ & ($0.11$) \\
            E-DBBC ($0.1$)  & $5.64$ & ($1.56$) & $18$ & $15.55$ & ($ 8.84$) & $0.90$ & ($0.05$) \\
            GE-DBBC ($0.1$) & $6.27$ & ($1.74$) & $18$ & $11.06$ & ($ 6.56$) & $0.90$ & ($0.05$) \\
            \hline          
        \end{tabular}
\end{table}

\clearpage

\section{Discussion}
\label{sec:conc}

In this paper, we developed model evaluation criteria that achieve the robustness in the variable selection and selection consistency simultaneously for the high-dimensional linear regression model based on statistical divergence measures. 
In order for robust and consistent selection of the regularization parameter to be realized, which divergence and penalty term we employ is crucial. 
Appropriate divergence with robustness, such as the BHHJ divergence, can diminish the damage caused by the contamination of outliers (Theorem \ref{theorem:robustness}), and precise approximation for the quasi-posterior probability built upon the divergence derives criteria that hold the selection consistency even when the number of candidate explanatory variables are quite large (Theorem \ref{theorem:consistency}). 
The adaptive LASSO-type regularization with appropriate penalty term that fulfils the conditions introduced in this paper leads the above-mentioned desirable characteristics in variable selection, without losing asymptotic properties such as the oracle property in estimation of regression coefficients. 
We also conducted various numerical examples to confirm the performance of our proposed criteria. 
The proposed criteria recorded high accuracy in comparison with non-robust criteria and methods without selection consistency, and the results agreed with the theorems shown in this paper. 

Noting that, there are various regularization methods other than the (adaptive) LASSO that we investigated thought this paper, for example, the elastic net (e.g., \cite{zou2005regularization}), $L^{q}$-penalty like the Bridge (e.g., \cite{frank1993statistical}), and $L^{0}$-penalty (e.g., \cite{bertsimas2016best, hastie2020best}). 
The study of robustify of such regularization methods will be productive, and comparing investigation of their properties will contribute higher accuracy of the variable selection for the high-dimensional regression. 
Moreover, although we focused on the asymptotic behavior and robustness against outliers for the normal linear regression model in this paper, our approach has a potential to apply to wider classes of models like the generalized linear model and functional data analysis. 
Additionally, theoretical and numerical results support the robustness of our proposed criteria against general contamination by outliers, yet we can expect more precise evaluation of robustness of the proposed criteria, such as exact distribution of estimators and criteria for contaminated populations, if we assume some specific contamination. 
These topics will be interesting and important subjects for future study.

\appendix

\section{On the regularity conditions}
\label{secappend:conditions}

\subsection{Conditions for proving the oracle property of the estimator and deriving the model evaluation criteria}
\label{secappend:conditions:Az}

We first introduce conditions on the optimization function of the estimation (Eq.~\eqref{eq:GALASSO:estimator} in Section \ref{sec:GALASSO}) and its associated terms, the regularization parameter, the explanatory variables, and the weights in the penalty term, to show the asymptotic properties of the (adaptive) LASSO-type estimators. 
Some of the conditions are related to the ones in \cite{fan2014adaptive}. 

\begin{assumption}
\label{assumption:estimation}
We assume the following conditions: 
\begin{description}
 \item[(A-1)] For each $i$ $=$ $1 \ldts n$, model's probability (density) function $f_{i}$ and the function $\rho_{i}$ in Eq.~\eqref{eq:GALASSO:estimator} are Lipschitz continuous, and each $\rho_{i}$ is three times continuously differentiable. 
 \item[(A-2)] With the regularization parameter $\lambda$ $=$ $\lambda_{n}$, $\lambda$ $>$ $2 \sqrt{\frac{(a_{1} + 1) \log P}{n}}$ holds for some positive constant $a_{1}$, and $s^{2} (\log n)^{\frac{5}{2}} n^{-\frac{3}{2}} \lambda^{-2}$ goes to $0$ as $n$ $\to$ $+\infty$. 
 \item[(A-3)] The maximum of absolute value in the design matrix $\bm{X}$ $=$ $(x_{i, j})_{i, j}$ is in constant-order $O(1)$. Moreover, the $s$ $\times$ $s$ matrix $\frac{1}{n} \, \bm{X}^{(1) \top} \bm{D}_{n} \, \bm{X}^{(1)}$ is positive definite, and all eigenvalues of this matrix are bounded below and above by some positive constants. Additionally, $\| \frac{1}{n} \, \bm{X}^{(2) \top} \bm{D}_{n} \, \bm{X}^{(1)} \|_{2, \infty}$ $<$ $a_{2} \, \lambda \sqrt{\frac{n}{s \, \log n}}$ $\min_{j \in \mathcal{J}^{(2)}} | w_{j} |$ holds for sufficiently large $n$ and a positive constant $a_{2}$, where $\| \bm{M} \|_{2,\infty}$ for a $p$ $\times$ $q$ matrix $\bm{M}$ is defined as $\| \bm{M} \|_{2,\infty}$ $=$ $\sup_{\bm{m} \in \breal^{q} \setminus \{ 0 \}} \frac{\| \bm{M} \, \bm{m} \|_{\infty}}{\| \bm{m} \|_{2}}$ ($p$ and $q$ are any positive integers). 
 \item[(A-4)] For the initial estimator $\hat{\bm{\beta}}^{\iniest}$ and a positive constant $a_{3}$, $\| \hat{\bm{\beta}}^{\iniest (1)} - \bm{\beta}_{\star}^{(1)} \|$ $\leq$ $a_{3} \sqrt{\frac{s \, \log P}{n}}$ holds with probability tending to one as $n$ $\to$ $+\infty$. 
 \item[(A-5)] With positive weight vector $\bm{w}$ $=$ $( \bm{w}^{(1) \top} ,\, \bm{w}^{(2) \top} )^{\top}$, $\lambda \| \bm{w}^{(1)} \|_{2}$ $\leq$ $a_{4} \sqrt{\frac{s}{n}}$ holds for a positive constant $a_{4}$ and for sufficiently large $n$, and there exists a positive constant $a_{5}$ such that $w_{j}$ $>$ $a_{5}$ for any $j$ $\in$ $\mathcal{J}^{(2)}$. Moreover, the weight function $h$ is non-increasing and Lipschitz continuous over $(0 ,\, +\infty)$, $h(a_{3} \sqrt{\frac{s \, \log P}{n}})$ $>$ $\frac{h(0+)}{2}$ holds for sufficiently large $n$, and $h'(z)$ $=$ $o(s^{-1} \lambda^{-1} n^{-\frac{1}{2}} (\log P)^{-\frac{1}{2}})$ holds for any $z$ $>$ $\frac{1}{2} \min_{j \in \mathcal{J}^{(1)}} |\beta^{\star}_{j}|$. 
 \item[(A-6)] The $s$ $\times$ $s$ matrix $\bm{Z}_{n}^{\top} \bm{\Omega}_{n} \, \bm{Z}_{n}$ is positive definite, and all eigenvalues of the matrices $\bm{V}_{n}$ and $\bm{Z}_{n}^{\top} \bm{\Omega}_{n} \, \bm{Z}_{n}$ are bounded below and above by some positive constants. 
 \item[(A-7)] Let $\bm{Q}_{n}$ $=$ $\ave \Bigl[ -\frac{\partial^{2} \sum_{i=1}^{n} \, \rho_{i}(Y_{i} - \bm{x}_{i}^{\top} \bm{\beta}_{\star})}{\partial \bm{\beta}^{(1)} \, \partial \bm{\beta}^{(1) \top}} \Bigr]$, then the value of the trace of $\bm{Q}_{n} \, \bm{U}_{n}^{-1} ( \bm{U}_{n}^{\top} )^{-1}$ is in between $O(1)$ and $O(s)$. 
\end{description}
\end{assumption}

In the normal linear regression models, we can confirm that (A-1) holds for the BHHJ divergence and most of the divergence measures in the JHHB and C families. 
(A-2) is the condition restricting the asymptotic order of the regularization parameter $\lambda$. 
The optimal order has also been discussed in many studies (e.g., \cite{fan2014adaptive, ghosh2024robust, zou2006adaptive}). 
(A-3) is the condition regarding the explanatory variables, that controls the correlation between the necessary and unnecessary variables. 
We can see that, the third condition in (A-3) immediately holds if the necessary and unnecessary explanatory variables are uncorrelated, i.e., $\bm{X}^{(2) \top} \bm{X}^{(1)}$ $=$ $\bm{O}_{(P-s) \times s}$ (see also \cite{fan2014adaptive}). 
(A-4) specifies the asymptotic behavior of the initial estimator $\hat{\bm{\beta}}^{\iniest}$. 
The LASSO-type estimator (solution of Eq.~\eqref{eq:GALASSO:estimator}) employing the BHHJ divergence and the uniform weight ($h(z)$ $=$ $1$ for any $z$ $>$ $0$) is an estimator satisfying this condition (see also \cite{ghosh2024robust}). 
(A-5) provides the asymptotic order of the weight function, and it supposes that the weights imposing on the necessary explanatory variables are asymptotically smaller than on the unnecessary ones. 
(A-6) is related to the condition in \cite{ghosh2024robust}, that is needed to obtain the asymptotic variance of the estimator. 
(A-7) is required to prove properties about the selection of the regularization parameter $\lambda$. 
We will provide a simple example when using the BHHJ divergence, in Subsection \ref{secappend:conditions:BHHJcase}.

\subsection{Conditions for establishing the selection consistency}
\label{secappend:conditions:Cz}

We next introduce conditions on the models and their associated terms, to prove the selection consistency (Theorem \ref{theorem:consistency}). 

\begin{assumption}
\label{assumption:consistency}
We assume the following conditions: 
\begin{description}
 \item[(C-1)] For any model $\mathcal{M}_{\iota}$ with $\nu(\iota)$ $\leq$ $S$, every eigenvalue of the $\nu(\iota)$ $\times$ $\nu(\iota)$ matrix $\frac{1}{n} \, \bm{H}_{M_{n}}(\bm{\beta}_{\star}(\iota))$ is positive and in constant-order $O(1)$, where $\bm{\beta}_{\star}(\iota)$ is a $\nu(\iota)$-dimensional vector defined as
 \begin{eqnarray}
  \bm{\beta}_{\star}(\iota) \:=\: \left( \left. \beta^{\star}_{j} \:\right|\: j \: \text{th explanatory variable is employed in model} \; \mathcal{M}_{\iota} \right) \nonumber
 \end{eqnarray}
  if $\mathcal{M}_{\iota}$ is an under-specified model, and $\bm{\beta}_{\star}(\iota)$ $=$ $( \beta_{j} )_{j}$ with 
 \begin{eqnarray}
  \beta_{j} \:=\: \left\{ \begin{array}{ll}
   \beta^{\star}_{j} & ( j \: \text{th explanatory variable is in the true model} \; \mathcal{M}_{\iota_{\star}} ) \\
   0 & (\text{otherwise}) \\
  \end{array} \right. \nonumber
 \end{eqnarray}
 for each $j$ if $\mathcal{M}_{\iota}$ is an over-specified model ($j$ $=$ $1 \ldts \nu(\iota)$). 
 \item[(C-2)] For any $\epsilon$ $>$ $0$, there exists a $\delta$ $>$ $0$ such that $(1 - \epsilon) \bm{H}_{M_{n}}(\bm{\beta}_{\star}(\iota))$ $\leq$ $\bm{H}_{M_{n}}(\bm{\beta}(\iota))$ $\leq$ $(1 + \epsilon) \bm{H}_{M_{n}}(\bm{\beta}_{\star}(\iota))$ for any model $\mathcal{M}_{\iota}$ with $\nu(\iota)$ $\leq$ $S$ and $\| \bm{\beta}(\iota) - \bm{\beta}_{\star}(\iota) \|_{2}$ $\leq$ $\delta$, when $n$ is sufficiently large. 
 \item[(C-3)] There exists a constant $\epsilon$ $>$ $0$ such that, for all $\delta$ $\in$ $(0 ,\, \epsilon)$, 
 \begin{eqnarray}
  \bm{\beta} \,\not \in\, \mathcal{B}_{\delta}(\hat{\bm{\beta}}(\iota_{\star}:\iota_{\mathrm{F}})) \;\Rightarrow\; \limsup_{n \to +\infty} \, \frac{1}{n} \left\{ R_{n}(\bm{\beta}) - R_{n}(\hat{\bm{\beta}}(\iota_{\star}:\iota_{\mathrm{F}})) \right\} \:<\: 0 \nonumber
 \end{eqnarray}
 holds, where $\hat{\bm{\beta}}(\iota:\iota_{\mathrm{F}})$ $=$ $( \hat{\beta}_{j}(\iota:\iota_{\mathrm{F}}) )_{j}$ for a model $\mathcal{M}_{\iota}$ is defined as
 \begin{eqnarray}
  \hat{\beta}_{j}(\iota:\iota_{\mathrm{F}}) \:=\: \left\{ \begin{array}{ll}
   \hat{\beta}^{\mathrm{F}}_{j} & ( j \: \text{th explanatory variable is in model} \; \mathcal{M}_{\iota} ) \\
   0 & (\text{otherwise}) \\
  \end{array} \right. \:, \nonumber
 \end{eqnarray}
 for each $j$, $\indexvec{\hat{\beta}^{\mathrm{F}}}{1}{P}$ is the estimator in the full model employing all of the $P$ explanatory variables, and $\mathcal{B}_{\delta}(\bm{\beta})$ indicates an open ball with center $\bm{\beta}$ and radius $\delta$. 
 \item[(C-4)] Every element in the $\nu(\iota)$ $\times$ $\nu(\iota)$ matrix $\bm{T}_{n}(\bm{\beta}(\iota))$ is bounded below and above by some positive constants in a neighborhood of $\bm{\beta}_{\star}(\iota)$ for any model $\mathcal{M}_{\iota}$. Moreover, there exist two positive constants $c_{1}$ and $c_{2}$ such that, $c_{1}$ $\leq$ $\bigl| \log | \bm{T}_{n}(\bm{\beta}(\iota)) | \bigr|$ $\leq$ $c_{2} \, \nu(\iota)$ holds in the neighborhood. 
\end{description}
\end{assumption}

(C-1) and (C-2) are expansions of the conditions given by \cite{chen2012extended}, that specify the behavior of the regression coefficients in the neighborhood of the true value. 
(C-3) is a sufficient condition for the assumptions described in \cite{kass1990validity} for conducting the Laplace approximation and establishing the selection consistency: it implies that the estimators based on under-specified models can not achieve asymptotically equivalent values of $R_{n}$ to the estimator based on the true model. 
This is a natural assumption, because, by definition, the distribution of an under-specified model never coincides with the true distribution no matter what values of parameters are used. 
(C-4) is the condition to restrict the asymptotic behavior of $\bm{T}_{n}$, that generally holds in the normal linear regression models when each value of explanatory variable is in constant-order.

\subsection{An example for the case of the BHHJ divergence}
\label{secappend:conditions:BHHJcase}

In this subsection, we give an example, in the case where $\frac{1}{n} \bm{X}^{\top} \bm{X}$ is equal to the identity matrix $\Imat_{P \times P}$ and every value in the design matrix $\bm{X}$ is in $O(1)$. 
We utilize the BHHJ divergence for the quasi-likelihood, and confirm the conditions (A-6) and (A-7) in Assumption \ref{assumption:estimation} in this case. 
As the distribution of $\epsilon_{i}$ $=$ $Y_{i}$ $-$ $\bm{x}_{i}^{\top} \bm{\beta}$ ($i$ $=$ $1 \ldts n$) under the model employing $\bm{\beta}$ as the regression coefficient vector is the normal distribution $\normaldist ( 0 ,\, \sigma^{2} )$, the function $\rho_{i}$ for each $i$ $=$ $1 \ldts n$ and its derivatives are calculated as 
\begin{eqnarray}
 \rho_{i}(\epsilon_{i}) &\bseq& \left( 2 \, \pi \right)^{-\frac{\alpha}{2}} \, \sigma^{-\alpha} \left\{ \, \left( \alpha + 1 \right)^{-\frac{1}{2}} - \frac{\alpha + 1}{\alpha} \, \exp \left( -\frac{\alpha \, \epsilon_{i}^{2}}{2 \, \sigma^{2}} \right) \right\} \:, \nonumber \\
 \rho_{i}'(\epsilon_{i}) &\bseq& \left( \alpha + 1 \right) \, \left( 2 \, \pi \right)^{-\frac{\alpha}{2}} \, \sigma^{-\alpha - 2} \, \epsilon_{i} \, \exp \left( -\frac{\alpha \, \epsilon_{i}^{2}}{2 \, \sigma^{2}} \right) \:, \nonumber \\
 \rho_{i}''(\epsilon_{i}) &\bseq& \left( \alpha + 1 \right) \, \left( 2 \, \pi \right)^{-\frac{\alpha}{2}} \, \sigma^{-\alpha - 2} \left( 1 - \frac{\alpha \, \epsilon_{i}^{2}}{\sigma^{2}} \right) \exp \left( -\frac{\alpha \, \epsilon_{i}^{2}}{2 \, \sigma^{2}} \right) \nonumber
\end{eqnarray}
for $\alpha$ $>$ $0$. 
We obtain the matrix $\bm{\Omega}_{n}$, the covariance matrix of $( \rho_{1}'(\epsilon_{1}) \ldts \rho_{n}'(\epsilon_{n}) )^{\top}$, and the diagonal matrix $\bm{D}_{n}$ $=$ $\diag ( \ave [ \rho_{1}''(\epsilon_{1}) ] \ldts \ave [ \rho_{n}''(\epsilon_{n}) ] )$ as 
\begin{eqnarray}
 \bm{\Omega}_{n} &\bseq& \left( \alpha + 1 \right)^{2} \, \left( 2 \, \alpha + 1 \right)^{-\frac{3}{2}} \, \left( 2 \, \pi \right)^{-\alpha} \, \sigma^{-2 \alpha - 2} \, \Imat_{n \times n} \:, \nonumber \\
 \bm{D}_{n} &\bseq& \left( \alpha + 1 \right)^{-\frac{1}{2}} \, \left( 2 \, \pi \right)^{-\frac{\alpha}{2}} \, \sigma^{- \alpha - 2} \, \Imat_{n \times n} \:, \nonumber
\end{eqnarray}
respectively. 
Clearly, elements in these matrices have values in constant-order with respect to $n$. 
As $\bm{X}^{(1) \top} \bm{X}^{(1)}$ $=$ $n \, \Imat_{s \times s}$, we can confirm that $\bm{V}_{n}$ $=$ $( \bm{X}^{(1) \top} \bm{D}_{n} \, \bm{X}^{(1)} )^{-\frac{1}{2}}$ $=$ $c \, n^{-\frac{1}{2}} \, \Imat_{s \times s}$ holds for a positive constant $c$ $=$ $O(1)$. 
Hereafter, we denote terms in constant-order by $c$ (this ``$c$'' does not always indicate the same value). 
Thus, $\bm{Z}_{n}$ $=$ $\bm{X}^{(1)} \, \bm{V}_{n}$ $=$ $c \, n^{-\frac{1}{2}} \, \bm{X}^{(1)}$, $\bm{Z}_{n}^{\top} \bm{\Omega}_{n} \, \bm{Z}_{n}$ $=$ $c \, \Imat_{s \times s}$, and $\bm{U}_{n}$ $=$ $( \bm{Z}_{n}^{\top} \bm{\Omega}_{n} \, \bm{Z}_{n} )^{-\frac{1}{2}} \bm{V}_{n}^{-1}$ $=$ $c \, n^{\frac{1}{2}} \, \Imat_{s \times s}$. 
We can see that $\bm{Z}_{n}^{\top} \bm{\Omega}_{n} \, \bm{Z}_{n}$ is clearly positive definite and the eigenvalues $\bm{V}_{n}$ and $\bm{Z}_{n}^{\top} \bm{\Omega}_{n} \, \bm{Z}_{n}$ are $O(1)$ due to $\max_{i, j} |x_{i, j}|$ $=$ $O(1)$, hence the condition (A-6) holds. 
Then, we confirm (A-7). 
We can calculate the matrices $\bm{Q}_{n}$ and $\bm{Q}_{n} \, \bm{U}_{n}^{-1} \left( \bm{U}_{n}^{\top} \right)^{-1}$ as follows: 
\begin{eqnarray}
 && \frac{\partial^{2} \sum_{i=1}^{n} \, \rho_{i}(Y_{i} - \bm{x}_{i}^{\top} \bm{\beta})}{\partial \bm{\beta}^{(1)} \, \partial \bm{\beta}^{(1) \top}} \:=\: \sum_{i=1}^{n} \, \rho_{i}''(Y_{i} - \bm{x}_{i}^{\top} \bm{\beta}) \, \bm{x}_{i}^{(1)} \, \bm{x}_{i}^{(1) \top} \:, \nonumber \\
 && \bm{Q}_{n} \:=\: \ave \left[ -\frac{\partial^{2} \sum_{i=1}^{n} \, \rho_{i}(Y_{i} - \bm{x}_{i}^{\top} \bm{\beta}_{\star})}{\partial \bm{\beta}^{(1)} \, \partial \bm{\beta}^{(1) \top}} \right] \:=\: -c \, n \, \Imat_{s \times s} \:, \nonumber \\
 && \bm{Q}_{n} \, \bm{U}_{n}^{-1} \left( \bm{U}_{n}^{\top} \right)^{-1} \:=\: - c \, \Imat_{s \times s} \:. \nonumber
\end{eqnarray}
Since the value of $\tr \bigl\{ \bm{Q}_{n} \, \bm{U}_{n}^{-1} ( \bm{U}_{n}^{\top} )^{-1} \bigr\}$ is proportional to $s$, the condition (A-7) holds.

\section{Proofs of theorems}
\label{secappend:proofs}

\begin{proof}[Proof of Theorem \ref{theorem:robustness}]
For proving this theorem, we use the following lemma provided by \cite{kurata2024robustness}. 

\begin{lemma}
\label{lemma:selectionrobustness}
For each $i$ $=$ $1 \ldts n$, let $f_{i}(\cdot\,|\,\bm{\beta})$, $g_{i}(\cdot)$, and $\xi_{i}(\cdot)$ be the probability (density) functions of the model distribution, the true distribution, and the outlier-generating distribution, respectively. 
When assuming the following conditions (R-1) and (R-2), the second-order approximation term of the difference of the value of $M_{n}$ (defined in Eq.~\eqref{eq:criteria:Mn}) built upon any divergence of the C divergence family with $\alpha$ $\geq$ $1$ between the case where the data-generating distribution is (non-contaminated) $\bm{G}$ and the case where that is (contaminated) $\bm{\Psi}$ is bounded for arbitrary outlier-generating distribution $\bm{\Xi}$. 
Moreover, for the BHHJ divergence family with $\alpha$ $>$ $0$ and JHHB divergence family with $\alpha$ $>$ $0$ and $\varphi > 0$, the boundedness holds under only (R-1). 
\begin{description}
\item[(R-1)] For all $i$, $\int f_{i}(y\,|\,\bm{\beta})^{\alpha}$ $\xi_{i}(y) \diff y$ and $\int \| \fracp{\log f_{i}(y\,|\,\bm{\beta})}{\bm{\beta}} \|$ $f_{i}(y\,|\,\bm{\beta})^{\alpha}$ $\xi_{i}(y) \diff y$ are finite for any $\bm{\beta}$ $\in$ $\Theta_{\star}$. 
\item[(R-2)] For all $i$, $\int f_{i}(y\,|\,\bm{\beta})^{\alpha - 1}$ $\xi_{i}(y)^{2} \diff y$, $\int f_{i}(y\,|\,\bm{\beta})^{\alpha - 1}$ $g_{i}(y)$ $\xi_{i}(y) \diff y$, and $\int \| \fracp{\log f_{i}(y\,|\,\bm{\beta})}{\bm{\beta}} \|$ $f_{i}(y\,|\,\bm{\beta})^{\alpha - 1}$ $g_{i}(y)$ $\xi_{i}(y) \diff y$ are finite for any $\bm{\beta}$ $\in$ $\Theta_{\star}$. 
\end{description}
\end{lemma}

The proof of this lemma is provided in the supplemental material in \cite{kurata2024robustness}. 
Incidentally, although ``$\alpha$ $>$ $1$'' is required for establishing the selection robustness of the C divergence-based criteria in general cases (\cite{kurata2024robustness}), we can also prove for $\alpha$ $=$ $1$ in the case of the normal linear regression models. 
Lemma \ref{lemma:selectionrobustness} supports the boundedness of second-order approximation term of the main term of the GE-DBBC against outliers. 
As the prior density in our setting and every term excepting the main term in the DBBC and E-DBBC (Eq.~\eqref{eq:criteria:EBICtype:def}) do not depend on the observation $\bm{Y}$, we can also confirm the boundedness of the values of the DBBC and E-DBBC. 
Hence, to prove the boundedness of the GE-DBBC (Eq.~\eqref{eq:criteria:GEBICtype:def}), we need to verify (R-1) and (R-2) and to check the behavior of $\log | \bm{T}_{n} |$ against outliers. 
Note that, in this proof, we omit the notation ``$(\iota)$'' in ``$\bm{\beta}(\iota)$'' for the sake of simplicity. 

In our setting, the probability density functions of $i$ th response variable in the model and true distributions are $f_{i}(y\,|\,\bm{\beta})$ $=$ $( 2 \, \pi \, \sigma^{2} )^{-\frac{1}{2}}$ $\exp \{ - \frac{(y - \bm{x}_{i}^{\top} \bm{\beta})^{2}}{2 \, \sigma^{2}} \}$ and $g_{i}(y)$ $=$ $f_{i}(y\,|\,\bm{\beta}_{\star})$, respectively. 
As the density $f_{i}$ is obviously bounded with respect to $y$, $\int f_{i}(y\,|\,\bm{\beta})^{\alpha}$ $\xi_{i}(y) \diff y$ in (R-1) is clearly finite for any outlier-generating distribution $\Xi_{i}$. 
The finiteness of $\int \| \fracp{\log f_{i}(y\,|\,\bm{\beta})}{\bm{\beta}} \|$ $f_{i}(y\,|\,\bm{\beta})^{\alpha}$ $\xi_{i}(y) \diff y$ also holds since $\fracp{\log f_{i}(y\,|\,\bm{\beta})}{\bm{\beta}}$ $=$ $\sigma^{-2}$ $( y - \bm{x}_{i}^{\top} \bm{\beta} )$ $\bm{x}_{i}$ and $m^{2} \exp (-\alpha \, m^{2})$ $\to$ $0$ as $m$ $\to$ $\pm \infty$ for any $\alpha$ $>$ $0$. 
Thus we can prove (R-1). 

Next, under assuming $\int \xi_{i}(y)^{2} \diff y$ $<$ $+\infty$, we can confirm the finiteness of $\int f_{i}(y\,|\,\bm{\beta})^{\alpha - 1}$ $\xi_{i}(y)^{2} \diff y$ for $\alpha$ $\geq$ $1$. 
Since $f_{i}(y\,|\,\bm{\beta})^{\alpha - 1}$ $g_{i}(y)$ $\propto$ $\exp \{ - \frac{(\alpha - 1) (y - \bm{x}_{i}^{\top} \bm{\beta})^{2}}{2 \, \sigma^{2}} \}$ $\exp \{ - \frac{(y - \bm{x}_{i}^{\top} \bm{\beta}_{\star})^{2}}{2 \, \sigma^{2}} \}$, $\int f_{i}(y\,|\,\bm{\beta})^{\alpha - 1}$ $g_{i}(y)$ $\xi_{i}(y) \diff y$ is finite for any $\alpha$ $\geq$ $1$. 
The finiteness of $\int \| \fracp{\log f_{i}(y\,|\,\bm{\beta})}{\bm{\beta}} \|$ $f_{i}(y\,|\,\bm{\beta})^{\alpha - 1}$ $g_{i}(y)$ $\xi_{i}(y) \diff y$ for $\alpha$ $\geq$ $1$ can be confirmed as with of $\int \| \fracp{\log f_{i}(y\,|\,\bm{\beta})}{\bm{\beta}} \|$ $f_{i}(y\,|\,\bm{\beta})^{\alpha}$ $\xi_{i}(y) \diff y$. 
Remark that, if $\alpha$ $<$ $1$, we can not support the finiteness of terms in (R-2). 

We finally consider the matrix $\bm{T}_{n}(\bm{\beta})$ $=$ $-\frac{1}{n} \fracppv{R_{n}(\bm{\beta})}{\bm{\beta}}$. 
In the setting of the normal linear regression model, when using the BHHJ divergence, the part depending on $\bm{\beta}$ of $R_{n}(\bm{\beta})$ $=$ $M_{n}(\bm{\beta})$ $+$ $\log \pi_{\iota}(\bm{\beta})$ is calculated as 
\begin{eqnarray}
 - \sum_{i=1}^{n} \, (2 \, \pi)^{-\frac{\alpha}{2}} \, \sigma^{-\alpha} \, \frac{\alpha + 1}{\alpha} \, \exp \left\{ - \frac{\alpha \, ( Y_{i} - \bm{x}_{i}^{\top} \bm{\beta} )^{2}}{2 \, \sigma^{2}} \right\} + n \, \lambda \sum_{j \in \mathcal{J}(\iota)^{(1)}} w_{j} \left| \beta_{j} \right| \:. \nonumber
\end{eqnarray}
Thus, by using $\fracp{| \beta_{j} |}{\beta_{j}}$ $=$ $\sign (\beta_{j})$ and $\fracpps{| \beta_{j} |}{\beta_{j}}$ $=$ $0$ for any $\beta_{j}$ $\neq$ $0$, we obtain the $\bm{T}_{n}(\bm{\beta})$ as follows: 
\begin{eqnarray}
 \bm{T}_{n}(\bm{\beta}) &\bseq& \frac{(\alpha + 1) \, (2 \, \pi)^{-\frac{\alpha}{2}} \, \sigma^{-\alpha - 2}}{n} \, \sum_{i=1}^{n} \, \left\{ 1 - \frac{\alpha \, ( Y_{i} - \bm{x}_{i}^{\top} \bm{\beta} )^{2}}{\sigma^{2}} \right\} \nonumber \\
 && \;\times\, \exp \left\{ - \frac{\alpha \, ( Y_{i} - \bm{x}_{i}^{\top} \bm{\beta} )^{2}}{\sigma^{2}} \right\} \, \bm{x}_{i} \, \bm{x}_{i}^{\top} \:. \nonumber
\end{eqnarray}
As the terms in the sum in the right-hand side of this equation are bounded for variation of the value of $Y_{i}$, the boundedness of $\log | \bm{T}_{n}(\bm{\beta}) |$ around $\bm{\beta}_{\star}$ holds due to (C-4) in Assumption \ref{assumption:consistency}. 
When using other divergence in the JHHB and C families, we can confirm that in a similar way. 
We complete the proof. 
\end{proof}

\begin{proof}[Proof of Theorem \ref{theorem:consistency}]
We here prove for only the GE-DBBC, since the proof for the E-DBBC can be conducted in a similar (or easier) way. 
Assume that $\mathcal{M}_{\iota}$ is a model employing $\nu(\iota)$ ($\leq$ $S$) explanatory variables and it is not coincide with the true model $\mathcal{M}_{\iota_{\star}}$. 
We will show the probability of the event $\{ \GEDBBC(\mathcal{M}_{\iota})$ $-$ $\GEDBBC(\mathcal{M}_{\iota_{\star}})$ $\leq$ $0 \}$, that is equal to $\{ B_{1}$ $-$ $B_{2}$ $\geq$ $0 \}$, tend to zero when $n$ goes to infinity, where 
\begin{eqnarray}
 B_{1} &\bseq& R_{n}(\hat{\bm{\beta}}(\iota)) - R_{n}(\hat{\bm{\beta}}(\iota_{\star})) \nonumber \\
 &\bseq& \left\{ M_{n}(\hat{\bm{\beta}}(\iota)) + \log \pi_{\iota}(\hat{\bm{\beta}}(\iota)) \right\} - \left\{ M_{n}(\hat{\bm{\beta}}(\iota_{\star})) + \log \pi_{\iota}(\hat{\bm{\beta}}(\iota_{\star})) \right\} \:, \nonumber \\
 B_{2} &\bseq& \left\{ \nu(\iota) - \nu(\iota_{\star}) \right\} \left\{ \frac{\log n - \log (2 \, \pi)}{2} + \left( 1 - \gamma \right) \, \log P \right\} \nonumber \\
 && \quad+\, \frac{1}{2} \left\{ \log \left| \bm{T}_{n}(\hat{\bm{\beta}}(\iota)) \right| - \log \left| \bm{T}_{n}(\hat{\bm{\beta}}(\iota_{\star})) \right| \right\} \:. \nonumber
\end{eqnarray}

\subsubsection*{(The case where $\mathcal{M}_{\iota}$ is an under-specified model)}

As the model $\mathcal{M}_{\iota}$ does not include the true model $\mathcal{M}_{\iota_{\star}}$, the ($L^{2}$)-norm $\| \hat{\bm{\beta}}(\iota:\iota_{\mathrm{F}})$ $-$ $\hat{\bm{\beta}}(\iota_{\star}:\iota_{\mathrm{F}}) \|_{2}$ is in $O_{P}(1)$ or more (at most $O_{P}(\sqrt{S})$), because every element in the true coefficient vector is in $O(1)$. 
Thus, $B_{1}$ takes a negative value almost surely and $B_{1}$ $=$ $O_{P}(n)$ when $n$ is sufficiently large, by (C-3) in Assumption \ref{assumption:consistency}. 
In addition, (C-4) in Assumption \ref{assumption:consistency} supports $\log | \bm{T}_{n}(\hat{\bm{\beta}}(\iota)) |$ $-$ $\log | \bm{T}_{n}(\hat{\bm{\beta}}(\iota_{\star})) |$ in $B_{2}$ is in between $O_{P}(1)$ and $O_{P}(S)$. 
For this reason, if $\nu(\iota)$ $=$ $\nu(\iota_{\star})$, $\prob \{ B_{1}$ $-$ $B_{2}$ $\geq$ $0 \}$ obviously goes to zero as $n$ $\to$ $+\infty$. 
Then, if $\nu(\iota)$ $<$ $\nu(\iota_{\star})$, the difference $\nu(\iota)$ $-$ $\nu(\iota_{\star})$ must be in the range $[-S ,\, -1]$, thus $B_{1}$ $\leq$ $-e_{1} \, n$ and $-B_{2}$ $\leq$ $e_{2} \, S \, \log P$ $=$ $e_{2} \, S \, n^{l}$ hold for sufficiently large $n$, $l \in (0 ,\, 1)$, and some positive constants $e_{1}$, $e_{2}$. 
Therefore, since $S$ $=$ $o( n^{\min \{ l / 2 ,\, (1 - l) / 2 \}} )$, $\prob \{ B_{1}$ $-$ $B_{2}$ $\geq$ $0 \}$ tends to zero as $n$ $\to$ $+\infty$. 
Moreover, if $\nu(\iota)$ $>$ $\nu(\iota_{\star})$, as $B_{2}$ takes positive value almost surely for sufficiently large $n$, $\prob \{ B_{1}$ $-$ $B_{2}$ $\geq$ $0 \}$ $\to$ $0$ ($n$ $\to$ $+\infty$) clearly holds. 
Hence, the probability of $\{ \GEDBBC(\mathcal{M}_{\iota})$ $-$ $\GEDBBC(\mathcal{M}_{\iota_{\star}})$ $\leq$ $0 \}$ tends to zero for any under-specified model $\mathcal{M}_{\iota}$ with $\nu(\iota)$ $\leq$ $S$. 

\subsubsection*{(The case where $\mathcal{M}_{\iota}$ is an over-specified model)}
We define the $s$ $=$ $\nu(\iota_{\star})$-dimensional vector $\bm{\beta}(\iota)^{(1)}$ for the $\nu(\iota)$-dimensional vector $\bm{\beta}(\iota)$. 
As an over-specified model includes the true model $\mathcal{M}_{\iota_{\star}}$, we can see that $\nu(\iota)$ $>$ $\nu(\iota_{\star})$ holds necessarily and both of $B_{1}$ and $B_{2}$ take positive values almost surely for sufficiently large $n$. 
Remark that, as every element in $\bm{\beta}(\iota)^{(1)}$ is non-zero around $\bm{\beta}_{\star}(\iota_{\star})$, $R_{n}(\bm{\beta}(\iota))$ is three times differentiable with respect to $\bm{\beta}(\iota)$ in a neighborhood of $\bm{\beta}_{\star}$ (see also (A-1) in Assumption \ref{assumption:estimation}). 
By the Taylor expansion, we obtain 
\begin{eqnarray}
 R_{n}(\hat{\bm{\beta}}(\iota)) - R_{n}(\bm{\beta}_{\star}(\iota)) &\bseq& \left\{ \hat{\bm{\beta}}(\iota)^{(1)} - \bm{\beta}_{\star}(\iota)^{(1)} \right\}^{\top} \check{\bm{q}}_{n}(\bm{\beta}_{\star}(\iota)) \nonumber \\
 && \quad+\, \frac{1}{2} \left\{ \hat{\bm{\beta}}(\iota)^{(1)} - \bm{\beta}_{\star}(\iota)^{(1)} \right\}^{\top} \check{\bm{Q}}_{n}(\bm{\beta}_{\star}(\iota)) \left\{ \hat{\bm{\beta}}(\iota)^{(1)} - \bm{\beta}_{\star}(\iota)^{(1)} \right\} \nonumber \\
 && \quad+\, O_{P} \left( \left\| \hat{\bm{\beta}}(\iota)^{(1)} - \bm{\beta}_{\star}(\iota)^{(1)} \right\|_{2}^{3} \right) \:, \label{eq:criteria:proof:overmodel:TF}
\end{eqnarray}
for sufficiently large $n$, where the $\nu(\iota)$-dimensional vector $\bm{\beta}_{\star}(\iota)$ was defined in Assumption \ref{assumption:consistency}, and $\check{\bm{q}}_{n}(\bm{\beta}(\iota))$ $=$ $\fracp{R_{n}(\bm{\beta}(\iota))}{\bm{\beta}(\iota)^{(1)}}$ and $\check{\bm{Q}}_{n}(\bm{\beta}(\iota))$ $=$ $\frac{\partial^{2} R_{n}(\bm{\beta}(\iota))}{\partial \bm{\beta}(\iota)^{(1)} \, \partial \bm{\beta}(\iota)^{(1) \top}}$. 
By ignoring the remainder term in Eq.~\eqref{eq:criteria:proof:overmodel:TF}, and using the completing the square and the Slutsky's theorem, we have 
\begin{eqnarray}
 R_{n}(\hat{\bm{\beta}}(\iota)) - R_{n}(\bm{\beta}_{\star}(\iota)) \:\simeq\: \frac{1}{2} \left( \bm{z} + \bm{Q}_{n}^{-1} \bm{q}_{n} \right)^{\top} \bm{Q}_{n} \, \left( \bm{z} + \bm{Q}_{n}^{-1} \bm{q}_{n} \right) - \frac{1}{2} \, \bm{q}_{n}^{\top} \bm{Q}_{n}^{-1} \bm{q}_{n} \:, \nonumber
\end{eqnarray}
where $\bm{z}$ $=$ $\hat{\bm{\beta}}(\iota)^{(1)}$ $-$ $\bm{\beta}_{\star}(\iota)^{(1)}$, $\bm{q}_{n}$ $=$ $\ave [ \check{\bm{q}}_{n}(\bm{\beta}_{\star}(\iota)) ]$, and $\bm{Q}_{n}$ $=$ $\ave [ \check{\bm{Q}}_{n}(\bm{\beta}_{\star}(\iota)) ]$. 
By Proposition \ref{proposition:estimation}, we can give an upper bound of the absolute value for sufficiently large $n$, as follows: 
\begin{eqnarray}
 \left| R_{n}(\hat{\bm{\beta}}(\iota)) - R_{n}(\bm{\beta}_{\star}(\iota)) \right| &\pbs \leq \pbs& \frac{1}{2} \, \left| \tr \left\{ \bm{Q}_{n} \, \bm{U}_{n}^{-1} \left( \bm{U}_{n}^{\top} \right)^{-1} \right\} \right| \,+\, \bm{q}_{n}^{\top} \bm{Q}_{n}^{-1} \bm{q}_{n} \:, \nonumber
\end{eqnarray}
By (A-7) in Assumption \ref{assumption:estimation}, the right-hand side of this equation is at most $O(S)$ for sufficiently large $n$, and thus, $B_{1}$ $\leq$ $e_{3} \, S$ holds for some positive constant $e_{3}$. 
On the other hand, as the difference $\nu(\iota)$ $-$ $\nu(\iota_{\star})$ is in the range $[1 ,\, S]$, the order of $B_{2}$ is in between $O(\log P)$ $=$ $O(n^{l})$ and $O(S \, n^{l})$. 
Since $S$ $=$ $o( n^{\min \{ l / 2 ,\, (1 - l) / 2 \}} )$, we can confirm that $B_{1}$ $-$ $B_{2}$ takes a negative value almost surely for sufficient large $n$, and hence the probability of $\{ \GEDBBC(\mathcal{M}_{\iota})$ $-$ $\GEDBBC(\mathcal{M}_{\iota_{\star}})$ $\leq$ $0 \}$ tends to zero for any over-specified model $\mathcal{M}_{\iota}$ with $\nu(\iota)$ $\leq$ $S$. 
We complete the proof. 
\end{proof}

\section{Details of the numerical simulations}
\label{secappend:tables}

We here provide the detailed results of the numerical simulations conducted in Subsection \ref{sec:numerical:simulation}. 
Tables \ref{tab:simulation:main:n5000P20}--\ref{tab:simulation:main:n200P1000} show the selection rates of the under-specified models (UM), true model (TM), and over-specified models (OM) for each pair of the sample size $n$ and the number of explanatory variables $P$. 

\begin{table}[ht]
    \caption{Selection rates (\%) of criteria for different contamination rates ($r$ $=$ $0$, $0.01$, $0.05$, and $0.1$) with $n=5000$ and $P=20$. Numerical values in parentheses indicate the values of the tuning parameter of the BHHJ divergence ($\alpha$): if $\alpha$ is equal to $0$, the corresponding criterion is based on the KL divergence, i.e., DBBC ($0$) and E-DBBC ($0$) are coincide with the BIC and EBIC, respectively. }
    \label{tab:simulation:main:n5000P20}
        \begin{tabular}{|l|rrr|rrr|rrr|rrr|} \hline
             & \multicolumn{3}{|c|}{$r = 0$} & \multicolumn{3}{|c|}{$r = 0.01$} & \multicolumn{3}{|c|}{$r = 0.05$} & \multicolumn{3}{|c|}{$r = 0.1$} \\ \hline 
            Criterion ($\alpha$) & UM & TM & OM & UM & TM & OM & UM & TM & OM & UM & TM & OM \\ \hline 
            Unif-BIC          & $  0$ & $ 95$ & $  5$ & $  0$ & $  5$ & $ 95$ & $  0$ & $  0$ & $100$ & $  0$ & $  0$ & $100$ \\
            Unif-EBIC         & $  0$ & $ 99$ & $  1$ & $  0$ & $  8$ & $ 92$ & $  0$ & $  0$ & $100$ & $  0$ & $  0$ & $100$ \\ \hline
            DBBC ($0$)        & $  0$ & $100$ & $  0$ & $  0$ & $ 96$ & $  4$ & $  0$ & $ 38$ & $ 62$ & $  0$ & $ 24$ & $ 76$ \\
            E-DBBC ($0$)      & $  0$ & $100$ & $  0$ & $  0$ & $ 98$ & $  2$ & $  0$ & $ 49$ & $ 51$ & $  0$ & $ 35$ & $ 65$ \\
            GE-DBBC ($0$)     & $  0$ & $100$ & $  0$ & $  0$ & $100$ & $  0$ & $  0$ & $ 99$ & $  1$ & $  0$ & $ 82$ & $ 18$ \\ \hline
            DBBC ($0.001$)    & $  0$ & $100$ & $  0$ & $  0$ & $100$ & $  0$ & $  0$ & $ 69$ & $ 31$ & $  0$ & $ 52$ & $ 48$ \\
            E-DBBC ($0.001$)  & $  0$ & $100$ & $  0$ & $  0$ & $100$ & $  0$ & $  0$ & $ 87$ & $ 13$ & $  0$ & $ 62$ & $ 38$ \\
            GE-DBBC ($0.001$) & $  0$ & $100$ & $  0$ & $  0$ & $100$ & $  0$ & $  0$ & $100$ & $  0$ & $  0$ & $100$ & $  0$ \\ \hline
            DBBC ($0.01$)     & $  0$ & $100$ & $  0$ & $  0$ & $100$ & $  0$ & $  0$ & $100$ & $  0$ & $  0$ & $100$ & $  0$ \\
            E-DBBC ($0.01$)   & $  0$ & $100$ & $  0$ & $  0$ & $100$ & $  0$ & $  0$ & $100$ & $  0$ & $  0$ & $100$ & $  0$ \\
            GE-DBBC ($0.01$)  & $  0$ & $100$ & $  0$ & $  0$ & $100$ & $  0$ & $  0$ & $100$ & $  0$ & $  0$ & $100$ & $  0$ \\ \hline
            DBBC ($0.1$)      & $  0$ & $100$ & $  0$ & $  0$ & $100$ & $  0$ & $  0$ & $100$ & $  0$ & $  0$ & $100$ & $  0$ \\
            E-DBBC ($0.1$)    & $  0$ & $100$ & $  0$ & $  0$ & $100$ & $  0$ & $  0$ & $100$ & $  0$ & $  0$ & $100$ & $  0$ \\
            GE-DBBC ($0.1$)   & $  0$ & $100$ & $  0$ & $  0$ & $100$ & $  0$ & $  0$ & $100$ & $  0$ & $  0$ & $100$ & $  0$ \\ \hline
            DBBC ($0.25$)     & $  0$ & $100$ & $  0$ & $  0$ & $100$ & $  0$ & $  0$ & $100$ & $  0$ & $  0$ & $100$ & $  0$ \\
            E-DBBC ($0.25$)   & $  0$ & $100$ & $  0$ & $  0$ & $100$ & $  0$ & $  0$ & $100$ & $  0$ & $  0$ & $100$ & $  0$ \\
            GE-DBBC ($0.25$)  & $  0$ & $100$ & $  0$ & $  0$ & $100$ & $  0$ & $  0$ & $100$ & $  0$ & $  0$ & $100$ & $  0$ \\ \hline
            DBBC ($0.5$)      & $  0$ & $100$ & $  0$ & $  0$ & $100$ & $  0$ & $  0$ & $100$ & $  0$ & $  0$ & $100$ & $  0$ \\
            E-DBBC ($0.5$)    & $  0$ & $100$ & $  0$ & $  0$ & $100$ & $  0$ & $  0$ & $100$ & $  0$ & $  0$ & $100$ & $  0$ \\
            GE-DBBC ($0.5$)   & $  0$ & $100$ & $  0$ & $  0$ & $100$ & $  0$ & $  0$ & $100$ & $  0$ & $  0$ & $100$ & $  0$ \\ \hline
            DBBC ($0.75$)     & $  0$ & $100$ & $  0$ & $  0$ & $100$ & $  0$ & $  0$ & $100$ & $  0$ & $  0$ & $100$ & $  0$ \\
            E-DBBC ($0.75$)   & $  0$ & $100$ & $  0$ & $  0$ & $100$ & $  0$ & $  0$ & $100$ & $  0$ & $  0$ & $100$ & $  0$ \\
            GE-DBBC ($0.75$)  & $  0$ & $100$ & $  0$ & $  0$ & $100$ & $  0$ & $  0$ & $100$ & $  0$ & $  0$ & $100$ & $  0$ \\ \hline
            DBBC ($1$)        & $  0$ & $100$ & $  0$ & $  0$ & $100$ & $  0$ & $  0$ & $100$ & $  0$ & $  0$ & $100$ & $  0$ \\
            E-DBBC ($1$)      & $  0$ & $100$ & $  0$ & $  0$ & $100$ & $  0$ & $  0$ & $100$ & $  0$ & $  0$ & $100$ & $  0$ \\
            GE-DBBC ($1$)     & $  0$ & $100$ & $  0$ & $  0$ & $100$ & $  0$ & $  0$ & $100$ & $  0$ & $  0$ & $100$ & $  0$ \\ \hline
        \end{tabular}
\end{table}

\begin{table}[ht]
    \caption{Results for $n=5000$ and $P=200$. }
    \label{tab:simulation:main:n5000P200}
        \begin{tabular}{|l|rrr|rrr|rrr|rrr|} \hline
             & \multicolumn{3}{|c|}{$r = 0$} & \multicolumn{3}{|c|}{$r = 0.01$} & \multicolumn{3}{|c|}{$r = 0.05$} & \multicolumn{3}{|c|}{$r = 0.1$} \\ \hline 
            Criterion ($\alpha$) & UM & TM & OM & UM & TM & OM & UM & TM & OM & UM & TM & OM \\ \hline 
            Unif-BIC          & $  0$ & $ 45$ & $ 55$ & $  0$ & $  1$ & $ 99$ & $  0$ & $  0$ & $100$ & $  0$ & $  0$ & $100$ \\
            Unif-EBIC         & $  0$ & $ 91$ & $  9$ & $  0$ & $  8$ & $ 92$ & $  0$ & $  0$ & $100$ & $  0$ & $  0$ & $100$ \\ \hline
            DBBC ($0$)        & $  0$ & $100$ & $  0$ & $  0$ & $ 80$ & $ 20$ & $  0$ & $ 18$ & $ 82$ & $  0$ & $ 14$ & $ 86$ \\
            E-DBBC ($0$)      & $  0$ & $100$ & $  0$ & $  0$ & $ 94$ & $  6$ & $  0$ & $ 39$ & $ 61$ & $  0$ & $ 19$ & $ 81$ \\
            GE-DBBC ($0$)     & $  0$ & $100$ & $  0$ & $  0$ & $100$ & $  0$ & $  0$ & $ 98$ & $  2$ & $  0$ & $ 47$ & $ 53$ \\ \hline
            DBBC ($0.001$)    & $  0$ & $100$ & $  0$ & $  0$ & $100$ & $  0$ & $  0$ & $ 57$ & $ 43$ & $  0$ & $ 21$ & $ 79$ \\
            E-DBBC ($0.001$)  & $  0$ & $100$ & $  0$ & $  0$ & $100$ & $  0$ & $  0$ & $ 79$ & $ 21$ & $  0$ & $ 31$ & $ 69$ \\
            GE-DBBC ($0.001$) & $  0$ & $100$ & $  0$ & $  0$ & $100$ & $  0$ & $  0$ & $100$ & $  0$ & $  0$ & $ 94$ & $  6$ \\ \hline
            DBBC ($0.01$)     & $  0$ & $100$ & $  0$ & $  0$ & $100$ & $  0$ & $  0$ & $100$ & $  0$ & $  0$ & $100$ & $  0$ \\
            E-DBBC ($0.01$)   & $  0$ & $100$ & $  0$ & $  0$ & $100$ & $  0$ & $  0$ & $100$ & $  0$ & $  0$ & $100$ & $  0$ \\
            GE-DBBC ($0.01$)  & $  0$ & $100$ & $  0$ & $  0$ & $100$ & $  0$ & $  0$ & $100$ & $  0$ & $  0$ & $100$ & $  0$ \\ \hline
            DBBC ($0.1$)      & $  0$ & $100$ & $  0$ & $  0$ & $100$ & $  0$ & $  0$ & $100$ & $  0$ & $  0$ & $100$ & $  0$ \\
            E-DBBC ($0.1$)    & $  0$ & $100$ & $  0$ & $  0$ & $100$ & $  0$ & $  0$ & $100$ & $  0$ & $  0$ & $100$ & $  0$ \\
            GE-DBBC ($0.1$)   & $  0$ & $100$ & $  0$ & $  0$ & $100$ & $  0$ & $  0$ & $100$ & $  0$ & $  0$ & $100$ & $  0$ \\ \hline
            DBBC ($0.25$)     & $  0$ & $100$ & $  0$ & $  0$ & $100$ & $  0$ & $  0$ & $100$ & $  0$ & $  0$ & $100$ & $  0$ \\
            E-DBBC ($0.25$)   & $  0$ & $100$ & $  0$ & $  0$ & $100$ & $  0$ & $  0$ & $100$ & $  0$ & $  0$ & $100$ & $  0$ \\
            GE-DBBC ($0.25$)  & $  0$ & $100$ & $  0$ & $  0$ & $100$ & $  0$ & $  0$ & $100$ & $  0$ & $  0$ & $100$ & $  0$ \\ \hline
            DBBC ($0.5$)      & $  0$ & $100$ & $  0$ & $  0$ & $100$ & $  0$ & $  0$ & $100$ & $  0$ & $  0$ & $100$ & $  0$ \\
            E-DBBC ($0.5$)    & $  0$ & $100$ & $  0$ & $  0$ & $100$ & $  0$ & $  0$ & $100$ & $  0$ & $  0$ & $100$ & $  0$ \\
            GE-DBBC ($0.5$)   & $  0$ & $100$ & $  0$ & $  0$ & $100$ & $  0$ & $  0$ & $100$ & $  0$ & $  0$ & $100$ & $  0$ \\ \hline
            DBBC ($0.75$)     & $  0$ & $100$ & $  0$ & $  0$ & $100$ & $  0$ & $  0$ & $100$ & $  0$ & $  0$ & $100$ & $  0$ \\
            E-DBBC ($0.75$)   & $  0$ & $100$ & $  0$ & $  0$ & $100$ & $  0$ & $  0$ & $100$ & $  0$ & $  0$ & $100$ & $  0$ \\
            GE-DBBC ($0.75$)  & $  0$ & $100$ & $  0$ & $  0$ & $100$ & $  0$ & $  0$ & $100$ & $  0$ & $  0$ & $100$ & $  0$ \\ \hline
            DBBC ($1$)        & $  0$ & $100$ & $  0$ & $  0$ & $100$ & $  0$ & $  0$ & $100$ & $  0$ & $  0$ & $100$ & $  0$ \\
            E-DBBC ($1$)      & $  0$ & $100$ & $  0$ & $  0$ & $100$ & $  0$ & $  0$ & $100$ & $  0$ & $  0$ & $100$ & $  0$ \\
            GE-DBBC ($1$)     & $  0$ & $100$ & $  0$ & $  0$ & $100$ & $  0$ & $  0$ & $100$ & $  0$ & $  0$ & $100$ & $  0$ \\ \hline
        \end{tabular}
\end{table}

\begin{table}[ht]
    \caption{Results for $n=500$ and $P=200$. }
    \label{tab:simulation:main:n500P200}
        \begin{tabular}{|l|rrr|rrr|rrr|rrr|} \hline
             & \multicolumn{3}{|c|}{$r = 0$} & \multicolumn{3}{|c|}{$r = 0.01$} & \multicolumn{3}{|c|}{$r = 0.05$} & \multicolumn{3}{|c|}{$r = 0.1$} \\ \hline 
            Criterion ($\alpha$) & UM & TM & OM & UM & TM & OM & UM & TM & OM & UM & TM & OM \\ \hline 
            Unif-BIC          & $  0$ & $ 36$ & $ 64$ & $  0$ & $  0$ & $100$ & $  0$ & $  0$ & $100$ & $  0$ & $  0$ & $100$ \\
            Unif-EBIC         & $  0$ & $ 69$ & $ 31$ & $  0$ & $  2$ & $ 98$ & $  0$ & $  0$ & $100$ & $  0$ & $  0$ & $100$ \\ \hline
            DBBC ($0$)        & $  0$ & $100$ & $  0$ & $  0$ & $ 36$ & $ 64$ & $  0$ & $  0$ & $100$ & $  0$ & $  0$ & $100$ \\
            E-DBBC ($0$)      & $  0$ & $100$ & $  0$ & $  0$ & $ 52$ & $ 48$ & $  0$ & $  0$ & $100$ & $  0$ & $  0$ & $100$ \\
            GE-DBBC ($0$)     & $  0$ & $100$ & $  0$ & $  0$ & $ 60$ & $ 40$ & $  0$ & $  0$ & $100$ & $  0$ & $  0$ & $100$ \\ \hline
            DBBC ($0.001$)    & $  0$ & $100$ & $  0$ & $  0$ & $ 92$ & $  8$ & $  0$ & $  0$ & $100$ & $  0$ & $  0$ & $100$ \\
            E-DBBC ($0.001$)  & $  0$ & $100$ & $  0$ & $  0$ & $ 97$ & $  3$ & $  0$ & $  4$ & $ 96$ & $  0$ & $  0$ & $100$ \\
            GE-DBBC ($0.001$) & $  0$ & $100$ & $  0$ & $  0$ & $ 99$ & $  1$ & $  0$ & $ 12$ & $ 88$ & $  0$ & $  0$ & $100$ \\ \hline
            DBBC ($0.01$)     & $  0$ & $100$ & $  0$ & $  0$ & $100$ & $  0$ & $  0$ & $ 98$ & $  2$ & $  0$ & $ 78$ & $ 22$ \\
            E-DBBC ($0.01$)   & $  0$ & $100$ & $  0$ & $  0$ & $100$ & $  0$ & $  0$ & $100$ & $  0$ & $  0$ & $ 92$ & $  8$ \\
            GE-DBBC ($0.01$)  & $  0$ & $100$ & $  0$ & $  0$ & $100$ & $  0$ & $  0$ & $100$ & $  0$ & $  0$ & $ 99$ & $  1$ \\ \hline
            DBBC ($0.1$)      & $  0$ & $100$ & $  0$ & $  0$ & $100$ & $  0$ & $  0$ & $100$ & $  0$ & $  0$ & $100$ & $  0$ \\
            E-DBBC ($0.1$)    & $  0$ & $100$ & $  0$ & $  0$ & $100$ & $  0$ & $  0$ & $100$ & $  0$ & $  0$ & $100$ & $  0$ \\
            GE-DBBC ($0.1$)   & $  0$ & $100$ & $  0$ & $  0$ & $100$ & $  0$ & $  0$ & $100$ & $  0$ & $  0$ & $100$ & $  0$ \\ \hline
            DBBC ($0.25$)     & $  0$ & $100$ & $  0$ & $  0$ & $100$ & $  0$ & $  0$ & $100$ & $  0$ & $  0$ & $100$ & $  0$ \\
            E-DBBC ($0.25$)   & $  0$ & $100$ & $  0$ & $  0$ & $100$ & $  0$ & $  0$ & $100$ & $  0$ & $  0$ & $100$ & $  0$ \\
            GE-DBBC ($0.25$)  & $  0$ & $100$ & $  0$ & $  0$ & $100$ & $  0$ & $  0$ & $100$ & $  0$ & $  0$ & $100$ & $  0$ \\ \hline
            DBBC ($0.5$)      & $  0$ & $100$ & $  0$ & $  0$ & $100$ & $  0$ & $  0$ & $100$ & $  0$ & $  0$ & $100$ & $  0$ \\
            E-DBBC ($0.5$)    & $  0$ & $100$ & $  0$ & $  0$ & $100$ & $  0$ & $  0$ & $100$ & $  0$ & $  0$ & $100$ & $  0$ \\
            GE-DBBC ($0.5$)   & $  0$ & $100$ & $  0$ & $  0$ & $100$ & $  0$ & $  0$ & $100$ & $  0$ & $  0$ & $100$ & $  0$ \\ \hline
            DBBC ($0.75$)     & $  0$ & $100$ & $  0$ & $  0$ & $100$ & $  0$ & $  0$ & $100$ & $  0$ & $  0$ & $100$ & $  0$ \\
            E-DBBC ($0.75$)   & $  0$ & $100$ & $  0$ & $  0$ & $100$ & $  0$ & $  0$ & $100$ & $  0$ & $  1$ & $ 99$ & $  0$ \\
            GE-DBBC ($0.75$)  & $  0$ & $100$ & $  0$ & $  0$ & $100$ & $  0$ & $  0$ & $100$ & $  0$ & $  1$ & $ 99$ & $  0$ \\ \hline
            DBBC ($1$)        & $  0$ & $100$ & $  0$ & $  0$ & $100$ & $  0$ & $  1$ & $ 99$ & $  0$ & $  0$ & $100$ & $  0$ \\
            E-DBBC ($1$)      & $  0$ & $100$ & $  0$ & $  0$ & $100$ & $  0$ & $  1$ & $ 99$ & $  0$ & $  0$ & $100$ & $  0$ \\
            GE-DBBC ($1$)     & $  2$ & $ 98$ & $  0$ & $  1$ & $ 99$ & $  0$ & $  4$ & $ 96$ & $  0$ & $  7$ & $ 93$ & $  0$ \\ \hline
        \end{tabular}
\end{table}

\begin{table}[ht]
    \caption{Results for $n=200$ and $P=1000$. }
    \label{tab:simulation:main:n200P1000}
        \begin{tabular}{|l|rrr|rrr|rrr|rrr|} \hline
             & \multicolumn{3}{|c|}{$r = 0$} & \multicolumn{3}{|c|}{$r = 0.01$} & \multicolumn{3}{|c|}{$r = 0.05$} & \multicolumn{3}{|c|}{$r = 0.1$} \\ \hline 
            Criterion ($\alpha$) & UM & TM & OM & UM & TM & OM & UM & TM & OM & UM & TM & OM \\ \hline 
            Unif-BIC          & $  0$ & $  4$ & $ 96$ & $  0$ & $  1$ & $ 99$ & $  0$ & $  0$ & $100$ & $  0$ & $  0$ & $100$ \\
            Unif-EBIC         & $  5$ & $ 31$ & $ 64$ & $  0$ & $  6$ & $ 94$ & $  0$ & $  0$ & $100$ & $  0$ & $  0$ & $100$ \\ \hline
            DBBC ($0$)        & $  0$ & $ 94$ & $  6$ & $  0$ & $ 14$ & $ 86$ & $  0$ & $  0$ & $100$ & $  0$ & $  0$ & $100$ \\
            E-DBBC ($0$)      & $  0$ & $ 99$ & $  1$ & $  0$ & $ 27$ & $ 73$ & $  0$ & $  0$ & $100$ & $  0$ & $  0$ & $100$ \\
            GE-DBBC ($0$)     & $  0$ & $ 98$ & $  2$ & $  0$ & $ 21$ & $ 79$ & $  0$ & $  0$ & $100$ & $  0$ & $  0$ & $100$ \\ \hline
            DBBC ($0.001$)    & $  0$ & $ 94$ & $  6$ & $  0$ & $ 18$ & $ 82$ & $  0$ & $  0$ & $100$ & $  0$ & $  0$ & $100$ \\
            E-DBBC ($0.001$)  & $  0$ & $ 99$ & $  1$ & $  0$ & $ 42$ & $ 58$ & $  0$ & $  0$ & $100$ & $  0$ & $  0$ & $100$ \\
            GE-DBBC ($0.001$) & $  0$ & $ 98$ & $  2$ & $  0$ & $ 25$ & $ 75$ & $  0$ & $  0$ & $100$ & $  0$ & $  0$ & $100$ \\ \hline
            DBBC ($0.01$)     & $  0$ & $ 92$ & $  8$ & $  0$ & $ 84$ & $ 16$ & $  0$ & $ 39$ & $ 61$ & $  0$ & $ 12$ & $ 88$ \\
            E-DBBC ($0.01$)   & $  2$ & $ 96$ & $  2$ & $  0$ & $ 99$ & $  1$ & $  0$ & $ 70$ & $ 30$ & $  1$ & $ 46$ & $ 53$ \\
            GE-DBBC ($0.01$)  & $  0$ & $ 98$ & $  2$ & $  0$ & $ 98$ & $  2$ & $  0$ & $ 70$ & $ 30$ & $  0$ & $ 42$ & $ 58$ \\ \hline
            DBBC ($0.1$)      & $  0$ & $ 87$ & $ 13$ & $  0$ & $ 86$ & $ 14$ & $  0$ & $ 85$ & $ 15$ & $  0$ & $ 85$ & $ 15$ \\
            E-DBBC ($0.1$)    & $  0$ & $ 98$ & $  2$ & $  0$ & $ 99$ & $  1$ & $  0$ & $ 98$ & $  2$ & $  0$ & $ 98$ & $  2$ \\
            GE-DBBC ($0.1$)   & $  0$ & $ 97$ & $  3$ & $  0$ & $ 99$ & $  1$ & $  0$ & $ 98$ & $  2$ & $  0$ & $ 96$ & $  4$ \\ \hline
            DBBC ($0.25$)     & $  0$ & $ 81$ & $ 19$ & $  0$ & $ 84$ & $ 16$ & $  0$ & $ 81$ & $ 19$ & $  0$ & $ 82$ & $ 18$ \\
            E-DBBC ($0.25$)   & $  1$ & $ 97$ & $  2$ & $  1$ & $ 97$ & $  2$ & $  5$ & $ 94$ & $  1$ & $  0$ & $ 97$ & $  3$ \\
            GE-DBBC ($0.25$)  & $  0$ & $ 98$ & $  2$ & $  0$ & $ 98$ & $  2$ & $  0$ & $ 99$ & $  1$ & $  0$ & $ 96$ & $  4$ \\ \hline
            DBBC ($0.5$)      & $  3$ & $ 90$ & $  7$ & $  4$ & $ 80$ & $ 16$ & $  3$ & $ 85$ & $ 12$ & $  2$ & $ 85$ & $ 13$ \\
            E-DBBC ($0.5$)    & $ 16$ & $ 84$ & $  0$ & $ 18$ & $ 82$ & $  0$ & $ 18$ & $ 81$ & $  1$ & $ 20$ & $ 78$ & $  2$ \\
            GE-DBBC ($0.5$)   & $  2$ & $ 97$ & $  1$ & $  2$ & $ 97$ & $  1$ & $  2$ & $ 97$ & $  1$ & $  3$ & $ 92$ & $  5$ \\ \hline
            DBBC ($0.75$)     & $  9$ & $ 89$ & $  2$ & $ 10$ & $ 83$ & $  7$ & $ 11$ & $ 76$ & $ 13$ & $ 14$ & $ 78$ & $  8$ \\
            E-DBBC ($0.75$)   & $ 39$ & $ 61$ & $  0$ & $ 52$ & $ 48$ & $  0$ & $ 45$ & $ 55$ & $  0$ & $ 43$ & $ 55$ & $  2$ \\
            GE-DBBC ($0.75$)  & $ 11$ & $ 89$ & $  0$ & $ 17$ & $ 82$ & $  1$ & $ 16$ & $ 82$ & $  2$ & $ 27$ & $ 72$ & $  1$ \\ \hline
            DBBC ($1$)        & $ 24$ & $ 76$ & $  0$ & $ 27$ & $ 70$ & $  3$ & $ 29$ & $ 67$ & $  4$ & $ 33$ & $ 62$ & $  5$ \\
            E-DBBC ($1$)      & $ 60$ & $ 40$ & $  0$ & $ 56$ & $ 44$ & $  0$ & $ 59$ & $ 40$ & $  1$ & $ 66$ & $ 33$ & $  1$ \\
            GE-DBBC ($1$)     & $ 42$ & $ 58$ & $  0$ & $ 40$ & $ 59$ & $  1$ & $ 49$ & $ 50$ & $  1$ & $ 58$ & $ 41$ & $  1$ \\ \hline
        \end{tabular}
\end{table}

\clearpage

\subsection*{Acknowledgements}

This work was supported by JSPS KAKENHI Grant Numbers JP20K19753, JP22H01139, JP23H00466, JP23H04474, JP23K01333, JP23K11007, JP25K21164, and JST-Mirai Program Grant Number JPMJMI18A2, Japan. 
R, a software environment for statistical computing and graphics (\cite{Rcoreteam}), was used for the data analysis.

\bibliographystyle{plain}
\bibliography{reference-v2}

\end{document}